%% file: at2019qiz_arxiv.tex
\documentclass[twocolumn,twocolappendix]{aastex63}
\usepackage{longtable,amsmath}

\newcommand{\kms}{km~s$^{-1}$}
\newcommand{\Msun}{M$_\sun$}

\newcommand{\swift}{\textsl{Swift}}
\defcitealias{2020mnRas.499..482N}{N20}
\submitjournal{ApJ}
\input{line_names}

\shorttitle{HST spectra of AT~2019qiz}
\shortauthors{Hung et al.}
\graphicspath{{./}{figures/}}

\begin{document}

\title{Discovery of a Fast Iron Low-Ionization Outflow in the Early Evolution of the Nearby Tidal Disruption Event AT~2019qiz}

\correspondingauthor{Tiara Hung}
\email{tiarahung@ucsc.edu}

\author[0000-0002-9878-7889]{Tiara Hung}
\affiliation{Department of Astronomy and Astrophysics,
University of California, Santa Cruz, CA, 95064, USA}

\author{Ryan~J.~Foley}
\affiliation{Department of Astronomy and Astrophysics,
University of California, Santa Cruz, CA 95064, USA}

\author{S. Veilleux}
\affiliation{Department of Astronomy, University of Maryland, College Park, MD  20742, USA}
\affiliation{Joint Space-Science Institute, University of Maryland, College Park, MD 20742, USA}

\author[0000-0003-1673-970X]{S.~B. Cenko}
\affiliation{Astrophysics Science Division, NASA Goddard Space Flight Center, MC 661, Greenbelt, MD 20771, USA}
\affiliation{Joint Space-Science Institute, University of Maryland, College Park, MD 20742, USA}

\author[0000-0002-9589-5235]{Jane~L.~Dai}
\affiliation{Department of Physics, University of Hong Kong, Pokfulam Road, Hong Kong}

\author{Katie~Auchettl}
\affiliation{School of Physics, The University of Melbourne, Parkville, VIC 3010, Australia}
\affiliation{ARC Centre of Excellence for All Sky Astrophysics in 3 Dimensions (ASTRO 3D), Australia}
\affiliation{Department of Astronomy and Astrophysics,
University of California, Santa Cruz, CA, 95064, USA}

\author[0000-0001-5955-2502]{Thomas G. Brink}
\affiliation{Department of Astronomy, University of California, Berkeley, CA 94720-3411, USA}

\author{Georgios~Dimitriadis}
\affiliation{Department of Astronomy and Astrophysics,
University of California, Santa Cruz, CA, 95064, USA}

\author[0000-0001-5955-2502]{Alexei V. Filippenko}
\affiliation{Department of Astronomy, University of California, Berkeley, CA 94720-3411, USA}
\affiliation{Miller Institute for Basic Research in Science, University of California, Berkeley, CA 94720, USA}

\author[0000-0003-3703-5154]{S. Gezari}
\affiliation{Department of Astronomy, University of Maryland, College Park, MD  20742, USA}
\affiliation{Joint Space-Science Institute, University of Maryland, College Park, MD 20742, USA}

\author[0000-0001-9206-3460]{Thomas~W.-S.~Holoien}
\altaffiliation{NHFP Einstein Fellow}
\affiliation{The Observatories of the Carnegie Institution for Science, 813 Santa Barbara Street, Pasadena, CA 91101, USA}

\author{Charles~D.~Kilpatrick}
\affiliation{Department of Astronomy and Astrophysics,
University of California, Santa Cruz, CA, 95064, USA}

\author{Brenna~Mockler}
\affiliation{Department of Astronomy and Astrophysics,
University of California, Santa Cruz, CA, 95064, USA}

\author{Anthony~L.~Piro}
\affiliation{The Observatories of the Carnegie Institution for Science, 813 Santa Barbara Street, Pasadena, CA 91101, USA}

\author{Enrico~Ramirez-Ruiz}
\affiliation{Department of Astronomy and Astrophysics,
University of California, Santa Cruz, CA, 95064, USA}
\affiliation{DARK, Niels Bohr Institute, University of Copenhagen, Lyngbyvej 2, 2100 Copenhagen, Denmark}

\author[0000-0002-7559-315X]{C\'{e}sar~Rojas-Bravo}
\affiliation{Department of Astronomy and Astrophysics,
University of California, Santa Cruz, CA, 95064, USA}

\author{Matthew~R.~Siebert}
\affiliation{Department of Astronomy and Astrophysics,
University of California, Santa Cruz, CA, 95064, USA}

\author[0000-0002-3859-8074]{Sjoert van Velzen}
\affiliation{Department of Astronomy, University of Maryland, College Park, MD 20742, USA}
\affiliation{Center for Cosmology and Particle Physics, New York University, NY 10003, USA}

\author[0000-0001-5955-2502]{WeiKang Zheng}
\affiliation{Department of Astronomy, University of California, Berkeley, CA 94720-3411, USA}

\begin{abstract}
We report the results of ultraviolet (UV) and optical photometric and spectroscopic analysis of the tidal disruption event (TDE) AT~2019qiz. Our follow-up observations started $<$10 days after the source began to brighten in the optical and lasted for a period of six months. Our late-time host-dominated spectrum indicates that the host galaxy likely harbors a weak active galactic nucleus. The initial {\it Hubble Space Telescope (HST)} spectrum of AT~2019qiz exhibits an iron and low-ionization broad absorption line (FeLoBAL) system that is seen for the first time in a TDE. This spectrum also bears a striking resemblance to that of Gaia16apd, a superluminous supernova. Our observations provide insights into the outflow properties in TDEs and show evidence for a connection between TDEs and engine-powered supernovae at early phases, as originally suggested by \cite{2016MNRAS.461..948M}. In a time frame of 50 days, the UV spectra of AT~2019qiz started to resemble those of previous TDEs with only high-ionization BALs. The change in UV spectral signatures is accompanied by a decrease in the outflow velocity, which began at $15,000$~\kms\ and decelerated to $\sim10,000$~\kms. A similar evolution in the \Ha\ emission-line width further supports the speculation that the broad Balmer emission lines are formed in TDE outflows. In addition, we detect narrow absorption features on top of the FeLoBAL signatures in the early {\it HST} UV spectrum of AT~2019qiz. The measured \ion{H}{1} column density corresponds to a Lyman-limit system, whereas the metal absorption lines (such as \ion{N}{5}, \ion{C}{4}, \ion{Fe}{2}, and \ion{Mg}{2}) are likely probing the circumnuclear gas and interstellar medium in the host galaxy.

\end{abstract}



\section{Introduction} \label{sec:intro}

Tidal disruption events (TDEs) refer to the transient phenomenon where a star on a close passage to a supermassive black hole (SMBH) is torn apart under tidal stress \citep{Hills1975}. For a star that initially traveled on a parabolic orbit, the disruption unbinds about half of the stellar mass while the bound other half assembles into an accretion disk and feeds the black hole until the debris streams are drained. In scenarios where disk formation is efficient, the onset of accretion resulting from a TDE is set by the fallback time ($t_{\rm fallback}$), which corresponds to the time it takes for the most bound debris on highly eccentric orbits to return to the pericenter \citep[e.g., $t_{\rm fallback} = 41\,M^{1/2}_\mathrm{BH,6}$ days, where $M_\mathrm{BH,6}$ is given in units of $10^6$\,M$_\odot$;][]{Lodato2009}. The liberated gravitational potential energy of the infalling gas is then converted to radiation across the electromagnetic spectrum, thus allowing a TDE to be detected by the observers \citep{1976MNRAS.176..633F,Rees1988}.

Observationally, X-ray, ultraviolet (UV), and optical sky surveys have identified nearly 5 dozen TDEs \cite[e.g.,][]{1999A&A...343..775K,1999A&A...349L..45K,2007A&A...462L..49E,Levan2011,Gezari2012,Holoien2016,2020arXiv200805461V}. However, these events, depending on the wavelength of discovery, exhibit a dichotomy in their properties. While the X-ray-detected TDEs are characterized by thermal emission that is consistent with the accretion model, most optically-detected TDEs appear to lack or have very weak ($\lesssim10^{-2}\,L_{\rm UV}$) X-ray emission \citep[e.g.,][]{2019ApJ...880..120H,2019ApJ...872..198V,2020arXiv200101409V}. In particular, the optically-detected events tend to have cooler temperatures that are 1--2 orders of magnitude lower than their X-ray counterparts \citep{vanVelzen2011,Arcavi2014,Holoien2014,2017ApJ...842...29H}. The optical TDEs are also able to maintain roughly the same temperature over a timescale of months.

The discrepancy between the two observational population of TDEs may arise from the biggest uncertainty in the current TDE framework, which is whether the stellar debris can circularize efficiently following a TDE. Circularization requires the bound debris in highly eccentric orbits to lose a large amount of orbital energy to form a circular disk at twice the pericenter radius, $2R_p$ (conservation of specific angular momentum). In the classical scheme, the energy is dissipated efficiently via shocks produced by self-intersection at the pericenter \citep{Evans1989,Rees1988,Phinney1989,2009ApJ...697L..77R}. This assumption is challenged by recent simulations and analytical calculations that find circularization to be extremely inefficient in certain regions of parameter space \citep{Dai15,2015ApJ...804...85S,2015ApJ...809..166G,2016MNRAS.461.3760H,2016MNRAS.455.2253B,2017MNRAS.tmp..118S}. This possibility lays the foundation for an alternative mechanism in which the UV and optical emission in TDEs is powered by the stream-stream collision shocks \citep{2015ApJ...804...85S,2015ApJ...806..164P}. However, the stream-stream collision model is radiatively inefficient \citep{2020arXiv200712198M}, which naturally leads to a slower light-curve evolution that is at odds with the observed $t^{-5/3}$ decline (theoretical mass fallback rate) in several TDEs \citep{2016MNRAS.461..948M}.

In TDEs where disk formation is efficient and where radiation is driven by the inner accretion flows, a reprocessing layer at 10--100\,$R_T$ (where $R_T$ is the tidal radius) is often invoked to explain the cooler temperature in optical TDEs. Some studies suggest that the bound debris could build up a hydrostatic envelope around the SMBH to reprocess the X-ray and extreme ultraviolet (EUV) radiation released by accretion \citep{1997ApJ...489..573L,Guillochon2013}, while others suggest that outflows could be the major reprocessing material in TDEs \citep{2015ApJ...805...83M,2016MNRAS.461..948M,2020ApJ...894....2P}. Indeed, radiation-driven winds are a natural consequence of TDEs given that a ``super-Eddington'' phase should be common among stellar disruptions by $M_\mathrm{BH}\lesssim10^7$\,\Msun\ black holes \citep{2009MNRAS.400.2070S,2018MNRAS.478.3016W}. For previous TDEs, the observed temperatures and luminosities generally correspond to an ejecta mass that is more massive than 1~\Msun\ \citep{2020arXiv200901240M}.

Multiwavelength follow-up observations of TDEs have detected winds across a wide range of velocities.
X-ray observations of ASASSN-14li revealed highly ionized outflows moving at both low and high velocities, from a few$\times$100~\kms\ to $0.2c$ \citep{2015Natur.526..542M,2018MNRAS.474.3593K}. The UV spectrum of ASASSN-14li also shows signs of a low-velocity outflow \citep{Cenko2016}, which has a velocity similar to that of the slower X-ray gas found by \cite{2015Natur.526..542M}. In the same event, radio observations are also supportive of the presence of either a subrelativistic outflow \citep{Alexander2016} or an off-axis relativistic jet \citep{vanVelzen2016}. Including AT~2019qiz, the subject of this paper, blueshifted broad absorption lines (BALs) that correspond to outflow velocities of 5000--15,000~\kms\ are detected in 4 out of 5 TDEs having {\it Hubble Space Telescope (HST)} UV follow-up spectroscopy, with the exception of ASASSN-14li \citep{2018MNRAS.473.1130B,2019ApJ...873...92B,2019ApJ...879..119H}. Orientation effects may explain why BALs are absent in some TDEs \citep{2020MNRAS.494.4914P}.

Contrary to the late-time observational properties of TDEs, which tend to be more uniform as the reprocessing layer becomes transparent and allows one to probe the accretion disk directly \citep{2020ApJ...889..166J,2018arXiv180900003V}, early-time observations are expected to exhibit a higher degree of diversity in the flare properties. Early-time multiwavelength observations of TDEs are critical for understanding debris-stream evolution and super-Eddington accretion in TDEs, though they have rarely been obtained. To date, there are only a handful of TDEs with published pre-peak multiband light curves \citep[e.g., ASASSN-18pg and ASASSN-19bt;][]{2020ApJ...898..161H,2019ApJ...883..111H}. Among these, ASASSN-19bt has the most densely-sampled rising optical light curve as it is located in the TESS Continuous Viewing Zone.

AT~2019qiz is the most well-observed TDE since ASASSN-19bt, with a wealth of early-time multiwavelength data. Here we present the analysis of the UV and optical data of AT~2019qiz, the current record holder of the nearest TDE at redshift $z=0.0151$. We note that \citet[][hereafter \citetalias{2020mnRas.499..482N}]{2020mnRas.499..482N} also analyzed the evolution of UV and optical broadband photometry and optical spectroscopy of AT~2019qiz. However, our analysis has a stronger focus on the unique multi-epoch {\it HST} UV spectra that have not been reported before. In addition, our $ugri$ light curves were obtained independently at a higher cadence than that of \citetalias{2020mnRas.499..482N} at early times.

This paper is structured as follows. The observations, including photometry and spectroscopy at UV and optical wavelengths, are discussed in \autoref{sec:obs}. We present the procedures and results of our analysis in \autoref{sec:analysis}. Specifically, in \autoref{sec:HST_UV_analysis} we detail the evolution and the identification of broad and narrow UV absorption lines in the {\it HST} spectra. We discuss in \autoref{sec:discussion} the implications of our results and the origin of the UV absorption lines. \autoref{sec:conclusions} presents our conclusions.

\section{Observations and Data Reduction}
\label{sec:obs}
AT~2019qiz is a fast-evolving nuclear transient in the nearby galaxy ($z=0.0151$) 2MASX J04463790-1013349. It was first discovered by the ATLAS \citep{2018PASP..130f4505T} on 2019 Sep. 18 (UT dates are used throughout this paper) in the orange filter and subsequently by the Zwicky Transient Facility \citep[ZTF;][]{2019PASP..131a8002B} on 2019 Sep. 19 in the $r$ filter; thus, it goes by the multiple names ATLAS19vfr and ZTF19abzrhgq. The most recent pre-flare nondetection was provided by ZTF on 2019 Sep. 16 with an upper limit of $g>19.22$~mag. We classified AT~2019qiz as a TDE from a Keck-I Low-Resolution Imaging Spectrometer \citep[LRIS;][]{1995PASP..107..375O} spectrum obtained on 2019 Sep. 25 \citep{2019TNSCR1921....1S} while still on the rise. Given the proximity of the TDE and the early classification, multiwavelength follow-up observations were triggered by the science community. Notably, AT~2019qiz was reported to be detected at radio frequencies and rising in 2019 October and November \citep[ATel \#13334;][]{2019ATel13334....1O}.

We detail the follow-up observations that are analyzed in this paper in the subsections below.
Throughout the paper, we adopt a flat $\Lambda$CDM cosmology with H$_0 = 67.4$~\kms~Mpc$^{-1}$ and $\Omega_m = 0.315$ measured by the {\it Planck} mission \citep{2020A&A...641A...6P}. The time difference ($\Delta t$) is expressed in rest-frame time with respect to the peak of $g$-band light curve at MJD 58763.93. All of the magnitudes are expressed in the AB system \citep{1983ApJ...266..713O}.

AT~2019qiz was simultaneously monitored by the Ultraviolet Optical Telescope \citep[UVOT;][]{gcg+04,2005SSRv..120...95R} and the X-ray Telescope (XRT) onboard the \textsl{Neil Gehrels Swift Observatory} \citep{gcg+04} during its flaring state. The X-ray emission from AT~2019qiz peaks at $\sim 10^{41}$~erg~s$^{-1}$ in the 0.3--10\,keV band, which is 2--3 orders of magnitude weaker than the UV and optical emission (\citetalias{2020mnRas.499..482N}). Given that the \swift\ XRT dataset has already been analyzed by \citetalias{2020mnRas.499..482N} and is independent from our other analysis, we reference their reported luminosity and hardness ratio where appropriate without repeating the data reduction and analysis in this work.

All data for AT~2019qiz have been corrected for Milky Way foreground extinction assuming a \citet{1989ApJ...345..245C} extinction curve with $R_{V} = 3.1$ and $E(B-V) = 0.0939 \pm 0.0088$~mag \citep{2011ApJ...737..103S}.

\subsection{HST STIS Spectra}
We obtained three epochs of UV spectra of AT~2019qiz with {\it HST} STIS (GO-16026; PI Hung) on 2019 Oct. 21, 2019 Dec. 12, and 2020 Jan. 15. The spectra were obtained through a 52\arcsec\ $\times$ 0\arcsec.2 aperture. For the near-UV (NUV) and far-UV (FUV) MAMA detectors, the G140L and G230L gratings were used to cover the spectral ranges of 1150--1730~$\textrm{\AA}$ and 1570--3180~$\textrm{\AA}$ at resolutions of 1.2~$\textrm{\AA}$ and 2.2~$\textrm{\AA}$, respectively. During the first two visits, the observation was obtained over a single {\it HST} orbit, with three equal exposures of 170~s in the NUV and three equal exposures of 335~s in the FUV. The observations obtained in the last visit consist of three 692~s exposures in the NUV and three 876~s exposures in the FUV, totaling two {\it HST} orbits. We used an inverse-variance weighting method to combine the 1-dimensional (1D) spectra at the same epoch that were output by the {\it HST} pipeline.

\subsection{Optical Photometry}
Following the classification of AT~2019qiz as a TDE on $\Delta t=-13$ days, we triggered photometric and spectroscopic monitoring spanning about 6 months in time (between 2019 Sep. and 2020 Mar.) before the TDE became too faint and Sun-constrained. \autoref{fig:lightcurve} shows the light curves of AT~2019qiz observed by the \swift\ UVOT and ground-based optical facilities including LCO, ZTF, and the Swope telescope at the Las Campanas Observatory. We list the photometric data in \autoref{tab: phot}.
Data reduction with each instrument is detailed in the following subsections.

\subsubsection{ZTF Photometry}
AT~2019qiz was simultaneously observed in the ZTF Mid-Scale Innovations Program field in both $g$ and $r$ with a 3-day cadence. The ZTF real-time pipeline performs standard image reduction and subtraction with respect to ZTF template images and distributes the events as alert packets on each observing night \citep{2019PASP..131a8002B,2019PASP..131a8003M}. We accessed the public alerts of AT~2019qiz via LCO MARS\footnote{\url{https://mars.lco.global}} and used the template-subtracted point-spread-function (PSF) magnitude to generate the light curves. The last ZTF detection was obtained on 2020 Feb. 25 in $g$ as ZTF stopped monitoring the field containing AT~2019qiz. We measured a signal-to-noise-ratio (S/N) weighted offset of 0.13\arcsec$\pm$0.17\arcsec\ from the ZTF $g$ and $r$ data, confirming that the transient is coincident with the galaxy nucleus, as expected for a TDE.

\subsubsection{Swope Photometry}
Optical photometry of AT~2019qiz in $ugri$ was obtained with the 1-m Swope telescope from 2019 Sep. 26 to 2020 Mar. 1 with a 2--5 day cadence. The images were reduced using the {\texttt photpipe} imaging and photometry pipeline \citep{Rest2005, Rest2014}. We subtracted the bias and flattened each frame using bias and sky-flat images obtained on the same night and in the same instrumental configuration as each AT~2019qiz image. The images were registered and geometric distortion was removed using 2MASS astrometric standards \citep{2MASS}. Using {\tt hotpants} \citep{HOTPANTS}, we subtracted pre-discovery Pan-STARRS1 (PS1) template images \citep{Flewelling16} from each Swope $gri$ frame.

Since we are not yet able to obtain a template image for the $u$ band, we instead extracted photometry within a 5\arcsec\ radius aperture and subtracted the host-galaxy light in $u$ by modeling the host emission. The $u$-band emission of the host galaxy is estimated by fitting the PS1 photometry in the $grizy$ bands and the 2MASS photometry in the $JHK$ filters in a 5\arcsec\ radius circular aperture with the synthetic stellar population fitting code \texttt{PROSPECTOR}. Our best-fit continuity star-formation history (SFH) model with 5 age bins yields a stellar mass of log$_{10}$($M_\mathrm{\star}$/\Msun) $= 10.43\pm0.04$ and a metal content of log$_{10}$($Z/Z_{\odot}$) = $-1.11^{+0.22}_{-0.45}$, which are consistent with values derived by \citetalias{2020mnRas.499..482N} and \cite{2020arXiv200101409V}.

\subsubsection{LCO Photometry}
We also obtained optical photometry of AT~2019qiz in the $ugr$ bands from 2019 Sep. 25 to 2019 Nov. 27 with the Sinistro camera mounted on one of the 1-m telescopes of the Las Cumbres Observatory (LCO) network in Siding Spring, Australia.
Similar to our handling of the Swope photometry, we removed host-galaxy contamination by performing image subtraction for the $g$ and $r$ bands and by modeling the host flux in the same 5\arcsec\ aperture for the $u$ band.

\subsection{UVOT Photometry}
We extracted UV light curves from a series of 39 \textsl{Swift} UVOT observations with the package \texttt{HEASoft} v6.27 and CALDB version 20200305. We estimated the counts of the source from a circular aperture of 5\arcsec\ radius and the background from a circular aperture of 40\arcsec\ radius using the task \texttt{UVOTSOURCE}. These were then converted to flux and magnitude with the \swift\ photometric calibration data \citep{2008MNRAS.383..627P,2011AIPC.1358..373B}.

\begin{figure*}
\centering
\includegraphics[width=7in, angle=0]{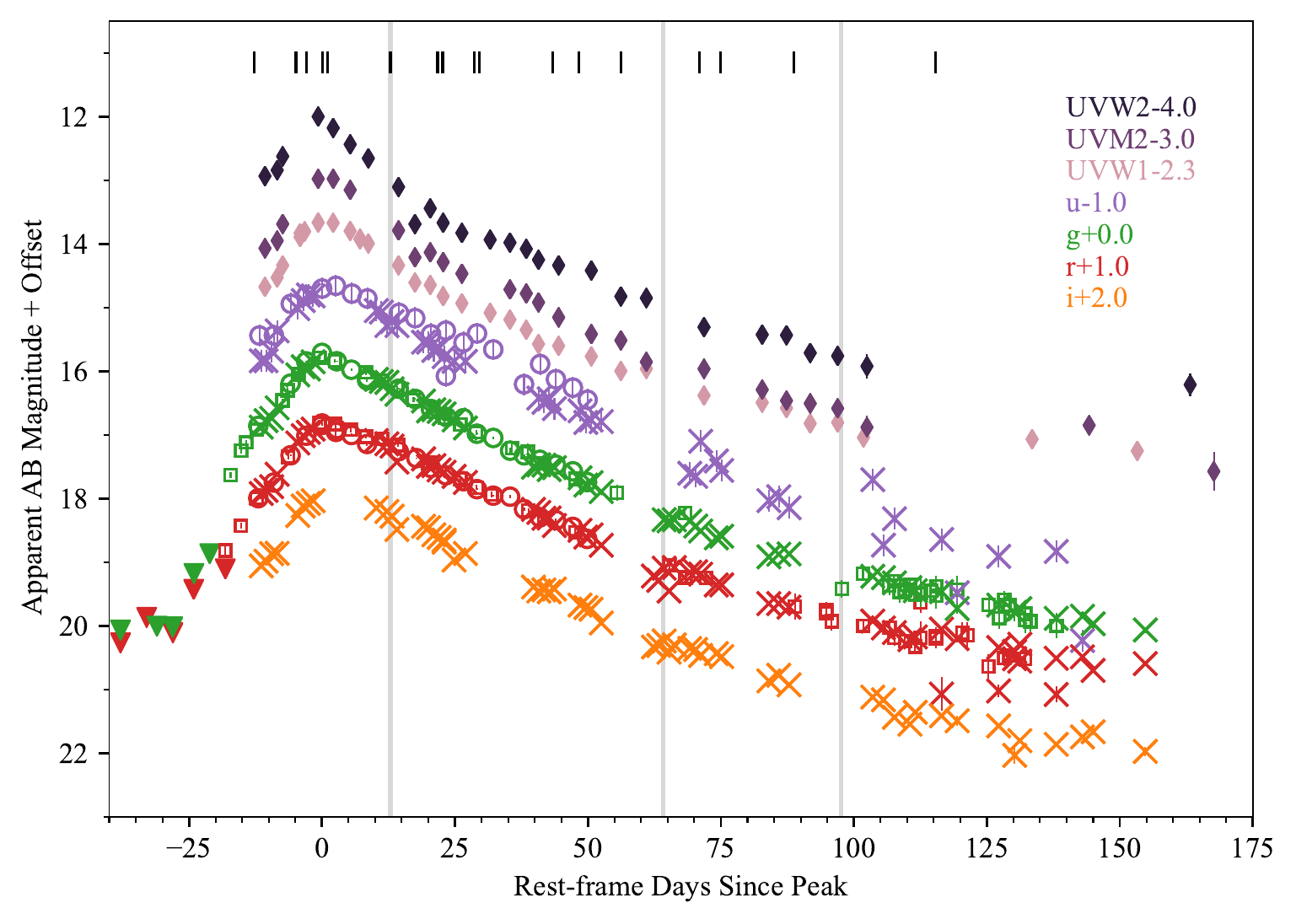}
\caption{Multiwavelength light curves of AT~2019qiz in AB magnitudes. Photometry obtained with \swift, ZTF, LCO, and Swope is marked with diamonds, squares, circles, and crosses, respectively. The triangles indicate the upper limits in the $g$ and $r$ filters. The light curves are color-coded by filters and offset by a constant to aid visualization. The host-galaxy contribution has been removed from all the photometry points obtained with ground-based optical telescopes. Although we did not subtract host-galaxy light from the \swift\ UV filters, the contamination is negligible compared to the transient light. We indicate the {\it HST} epochs with gray vertical lines, whereas the short black vertical lines mark the epochs with optical spectroscopic observations.
  \label{fig:lightcurve}}
\end{figure*}

The \swift\ target-of-opportunity (ToO) observations covered the evolution of AT~2019qiz from $\Delta t=-11$ to $168$ days. Although the observations were made in all six UVOT filters (UVW2, UVM2, UVW1, $U$, $B$, and $V$), in \autoref{fig:lightcurve} we only show the data from the three bluest filters (UVW2, UVM2, and UVW1), where the host-galaxy contributions are negligible. These are also the only \swift\ filters used in our data analysis. The nondetections in the archival {\it GALEX} All-Sky Imaging Survey place an upper limit of FUV $>20$ mag and NUV $>20.8$ mag on the host galaxy light.  Given that the flux in the \swift\ UV filters is highly dominated by the TDE at the time of the observations, we did not attempt subtracting the host-galaxy flux from these bands.

\subsection{Optical Spectroscopy}

We obtained a total of 18 spectroscopic observations with the Kast spectrograph \citep{Miller1993} on the Lick 3-m Shane telescope, the Goodman spectrograph on the SOAR telescope \citep{2004SPIE.5492..331C}, and LRIS on the Keck~I 10-m telescope.
Detailed instrumental configurations are listed in \autoref{tab:obs_spec}. We performed 1D spectrum extraction and flux calibration with standard \texttt{PyRAF}\footnote{\url{http://www.stsci.edu/institute/software_hardware/pyraf}} routines. Observations of standard stars BD+174708 and BD+284211 were used to determine the relative flux calibration and remove telluric features \citep[e.g.,][]{foley03, silverman12, dimitriadis19}. All of the spectra presented in this paper have been corrected for Galactic extinction.
We calibrated each spectrum's absolute flux by comparing its $g$-band synthetic photometry to the photometry from Swope and LCO imaging data (including host contribution), interpolated to each spectroscopic epoch.

\begin{table*}
\centering
\caption{Observation details of the optical spectra of AT~2019qiz}
\label{tab:obs_spec}
\bigskip
\begin{tabular}{lccccc}
\hline
\hline
Obs. Date & Phase (days) & Telescope + Instrument & Slit Width & Grism/Grating & Exp. Time (s) \\
\hline
2019-09-25    &   $-13$     &   Keck + LRIS      &  1.0\arcsec  &  600/400 + 400/8500  &  450 (blue) 438 (red) \\
2019-10-03    &   $-5$      &   Shane + Kast     &  2.0\arcsec  &  452/3306 + 300/7500 &  3660 (blue) 3600 (red)\\
2019-10-05    &   $-3$      &   Shane + Kast     &  2.0\arcsec  &  452/3306 + 300/7500 &  930 (blue) 2$\times$450 (red) \\
2019-10-08    &   0       &   Shane + Kast     &  2.0\arcsec  &  452/3306 + 300/7500 &  1560 (blue) 1500 (red) \\
2019-10-09    &   1       &   SOAR + Goodman   &  1.0\arcsec  &  400 m2              &  720 \\
2019-10-21    &   13      &   SOAR + Goodman   &  1.0\arcsec  &  400 m2              &  2$\times$720  \\
2019-10-30    &   22      &   Shane + Kast     &  2.0\arcsec  &  452/3306 + 300/7500 &  1230 (blue) 2$\times$600 (red)\\
2019-10-31    &   23      &   Shane + Kast     &  2.0\arcsec  &  452/3306 + 300/7500 &  1560 (blue) 1500 (red) \\
2019-11-06    &   29      &   Shane + Kast     &  2.0\arcsec  &  452/3306 + 300/7500 &  1865 (blue) 3$\times$600 (red) \\
2019-11-07    &   30      &   Shane + Kast     &  2.0\arcsec  &  300/7500            &  1500 (red) \\
2019-11-21    &   43      &   Shane + Kast     &  2.0\arcsec  &  300/7500            &  2400 (red) \\
2019-11-26    &   48      &   Keck + LRIS      &  1.0\arcsec  &  600/400 + 400/8500  &  450 (blue) 438 (red)  \\
2019-12-04    &   56      &   SOAR + Goodman   &  1.0\arcsec  &  400 m2              &  1200   \\
2019-12-19    &   71      &   SOAR + Goodman   &  1.0\arcsec  &  400 m2              &  2$\times$1320  \\
2019-12-23    &   75      &   Shane + Kast     &  2.0\arcsec  &  452/3306 + 300/7500 &  2$\times$1230 (blue) 4x600 (red)  \\
2020-01-06    &   89      &   Shane + Kast     &  2.0\arcsec  &  452/3306 + 300/7500 &  1845 (blue) 3$\times$600 (red) \\
2020-02-02    &   115     &   Shane + Kast     &  2.0\arcsec  &  452/3306 + 300/7500 &  1230 (blue) 2$\times$600 (red) \\
2020-09-18    &   340     &   Keck + LRIS      &  1.0\arcsec  &  600/400 + 400/8500  &  910 (blue) 900 (red)  \\
\hline
\hline
\end{tabular}
\end{table*}

\section{ANALYSIS} \label{sec:analysis}

\subsection{Black Hole Mass Estimation}\label{subsec:bh_mass}

We have not yet been able to obtain high-resolution spectra to measure the stellar velocity dispersion of the host galaxy. Nevertheless, \citetalias{2020mnRas.499..482N} measured a velocity dispersion of $\sigma=69.7\pm2.3$\,\kms\ from their late-time X-shooter spectrum. Using the scaling relation derived for a sample of low-mass galaxies from \cite{2011ApJ...739...28X}, this velocity dispersion corresponds to a black hole mass of log$_{10}(M_\mathrm{BH}$/\Msun) = ${6.16\pm0.43}$. Different $M-\sigma$ relations generally agree with an upper limit of log$_{10}(M_\mathrm{BH}$/\Msun) $\lesssim 6.5$ \citep{2020mnRas.499..482N}. As detailed below, the black hole mass derived from light-curve fitting is also consistent with this value. Therefore we use $M_\mathrm{BH} = 1.4\times10^6$\,\Msun\ when estimating relevant scales throughout the paper.

We also used the python package Modular Open-Source Fitter for Transients \citep[\texttt{MOSFiT};][]{2018ApJS..236....6G} to simulate the observed UV and optical light curves in \autoref{fig:lightcurve} to derive the physical parameters of the TDE, including the black hole mass. The TDE model implemented in \texttt{MOSFiT} estimates the bolometric luminosity of the TDE by converting the mass fallback rates from hydrodynamic simulations \citep{Guillochon2013} with a constant efficiency parameter \citep{2018ApJS..236....6G,Mockler2019}. The program then tries to match the observed flux in each band by reprocessing the bolometric flux with a blackbody photosphere, assuming that the blackbody photosphere evolves as a power law of the mass-fallback rate.

The TDE model in \texttt{MOSFiT} has 8 parameters: black hole mass ($M_{\rm h}$), stellar mass ($M_{\ast}$), scaled impact parameter ($b$), photosphere power-law exponent ($l$), photosphere radius normalization constant ($R_{\rm ph0}$), efficiency ($\epsilon$), viscous delay time ($t_{\rm viscous}$), and the fallback time of the most bound debris ($t_\mathrm{fallback}$). We present the best-fit values from the \texttt{MOSFiT} run for AT~2019qiz in \autoref{tab:mosfit}. The only constraint we imposed on the fitting parameters is for the stellar mass ($M_*$) to stay below 3\,\Msun. This is because $M_*$ is not always well-constrained by \texttt{MOSFiT} owing to its degeneracy with the efficiency parameter, and it is physically unlikely for $M_*$ to exceed 3\,\Msun\ since higher mass stars tend to have shorter lifetimes.

Our derived parameters are generally consistent with those of \citetalias{2020mnRas.499..482N}, though the fitting was performed on a different set of optical data. Our results suggest a negligible viscous time, therefore $t_\mathrm{fallback}$ approximates the start of the flare. We derived $t_\mathrm{fallback} = -25$ days relative to the time of peak light while \citetalias{2020mnRas.499..482N} found $t_\mathrm{fallback} = -27$ days. This value is also consistent with that estimated from a photosphere expanding at constant velocity, where $\Delta t=-31\pm2$ days (see \autoref{sec:sed_fit}). The black hole mass of log$_{10}(M_\mathrm{BH}$/\Msun) $= {6.14\pm0.1}$ derived from \texttt{MOSFiT} is in good agreement with that estimated from the $M-\sigma$ relation.

\begin{deluxetable}{lc}
\centering
\tablecaption{\texttt{MOSFiT} best-fit parameters\label{tab:mosfit}}
\tablehead{\colhead{Parameter$^a$} & \colhead{Value}}
\startdata
$t_{\rm fallback}$ (days)   & $-25_{-1}^{+1}$        \\
log$_{10} R_{ph0}$               & $0.9_{-0.09}^{+0.06}$     \\
log$_{10} t_{\rm viscous}$ (days)    & $0.61_{-0.29}^{+0.12}$    \\
$l_{\rm photo}$                      & $0.66_{-0.03}^{+0.03}$      \\
$\beta$                              & $0.58_{-0.03}^{+0.03}$     \\
log$_{10} M_{\rm h}$ (M$_{\odot}$) & $6.14_{-0.1}^{+0.09}$     \\
log$_{10} \epsilon$      & $-3.47_{-0.23}^{+0.13}$            \\
$M_{\ast}$ (M$_{\odot}$)               & $1.1_{-0.1}^{+0.97}$     \\
\enddata
\tablecomments{$^a$ See \autoref{subsec:bh_mass} for parameter definition and \cite{Mockler2019} for detailed model description.}
\end{deluxetable}

\subsection{Blackbody Temperature, Radius, and Bolometric Luminosity Evolution} \label{sec:sed_fit}
\begin{figure}
\centering
\includegraphics[width=3.5in, angle=0]{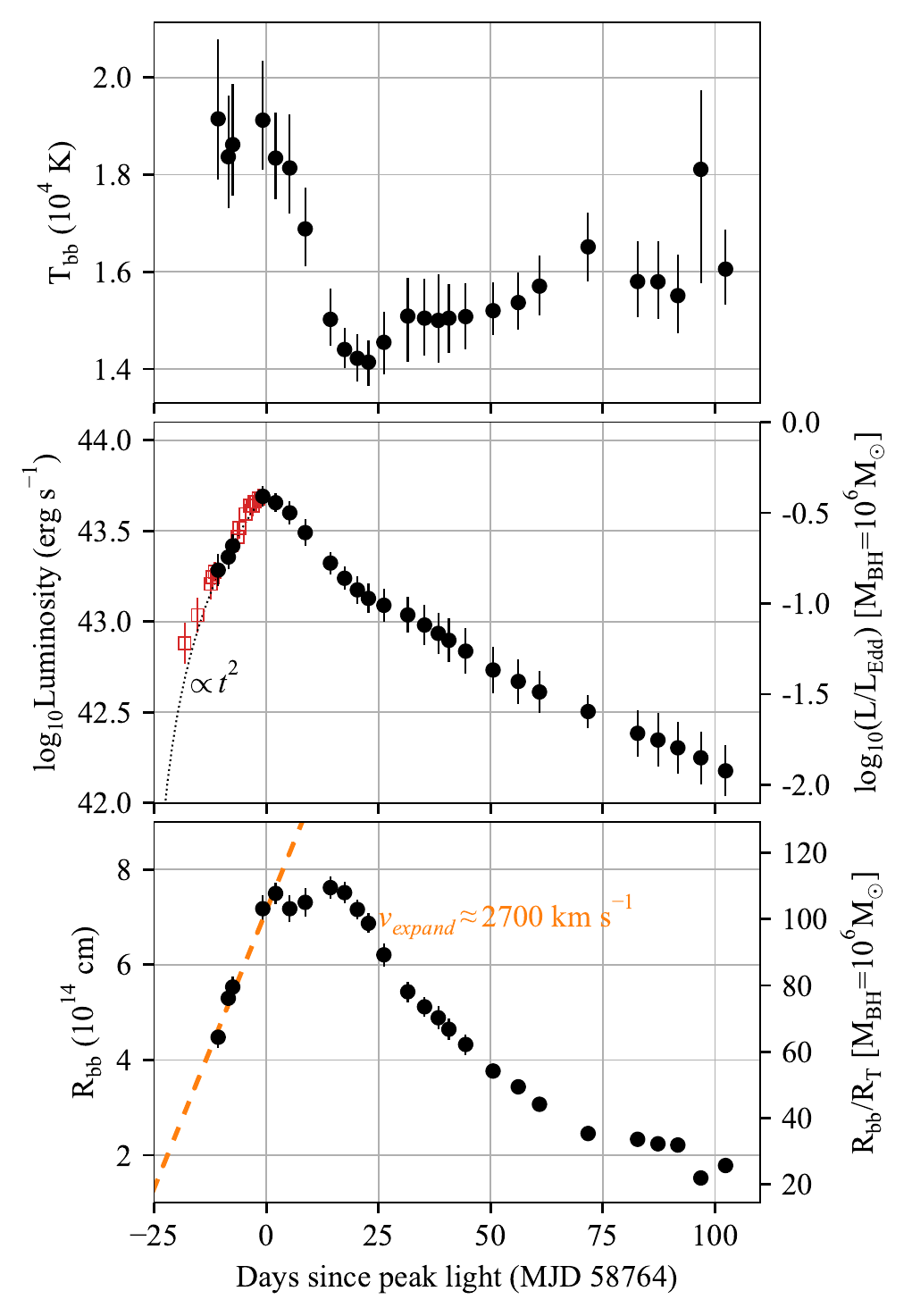}
\caption{The evolution of blackbody temperature (top), bolometric luminosity (middle), and blackbody radius (bottom) measured from the UV and optical light curves at \swift\ epochs (black circles). We also show the luminosity scaled from the $r$-band light curve (open red squares) for the rising part to help visualize the early evolution. The dotted line in the middle panel shows the best-fit quadratic function to the $L_{\rm bol}$ measured from the \swift\ epochs during the rise. The quadratic function asymptotically approaches a reference time $t_0 = -26.2$ days. The dashed orange curve displays the best-fit expansion velocity to the rising part of $R_{\rm bb}$.
  \label{fig:bbparams_evo}}
\end{figure}

As a standard TDE analysis procedure, we model the spectral energy distribution (SED) of AT~2019qiz with a single-temperature blackbody. To do so, we first construct the SEDs at epochs having \swift\ observations in all three UV filters ($UVW2$, $UVM2$, and $UVW1$) and interpolate the photometry in the $ugri$ filters measured from ground-based observatories.
We show the best-fit blackbody temperature ($T_{\rm bb}$) and the 10\%--90\% confidence level in \autoref{fig:bbparams_evo}. We derive the bolometric luminosity ($L_{\rm bol}$) at each \swift\ epoch by integrating over the best-fit blackbody spectra and calculate the emitting radius of the blackbody with the Stefan-Boltzmann law, where $L_{\rm bol}=4\pi R_{\rm bb}^2\sigma T_{\rm bb}^4$. The evolution of $L_{\rm bol}$ and $R_{\rm bb}$ are also given in \autoref{fig:bbparams_evo}. Our measurements are in good agreement with those of \citetalias{2020mnRas.499..482N}.

The blackbody temperature of AT~2019qiz initially had a constant value of $T_{\rm bb}\approx1.9\times10^4$~K as the light curves approached maximum brightness ($t_{\rm peak}$). From day 0 to day 25, it started to cool significantly down to $T_{\rm bb}\approx1.4\times10^4$~K. Afterward, $T_{\rm bb}$ slowly recovered back to $\sim 1.6\times10^4$~K and remained roughly constant out to day 100. This initial decline in $T_{\rm bb}$ was also observed in ASASSN-14ae \citep{Holoien2014} shortly after discovery around peak light, and in ASASSN-19bt while the light curves were rising \citep{2019ApJ...883..111H}. This cooing phase is short enough to be missed by previous TDE discoveries, where the classification and UV follow-up observations typically came around or after maximum light; hence, only the constant-temperature phase was observed.

Our derived peak luminosity, $L_{\rm peak}=4.9\times10^{43}$ erg s$^{-1}$, corresponds to an Eddington ratio of 0.27 assuming a black hole mass of $1.4\times10^6$\,\Msun\ estimated from the $M-\sigma$ relation. We determine a total radiated energy of $1.3\times10^{50}$ erg by extrapolating the luminosity linearly out to $t=\pm\infty$. If entirely powered by accretion, this energy would imply an accreted mass of $6.9\times10^{-4}(\epsilon/0.1)^{-1}$\,\Msun, where $\epsilon$ is the accretion efficiency. The small accreted mass compared to the bound stellar mass could imply a low radiative efficiency or a partial disruption, or that we are underestimating the TDE energetics in other wavelengths (e.g., EUV or infrared). In fact, we see indications that the best-fit blackbody tends to underestimate the FUV continuum in our \textit{HST} spectra.
The luminosity evolution of AT~2019qiz places it in an unoccupied strip on the luminosity-phase plot that is between the ``fast and faint'' TDE iPTF16fnl and the slower but brighter TDE population (see Fig. 1 of \cite{2020arXiv200805461V} and Fig. 10 of \citetalias{2020mnRas.499..482N}).

The blackbody radius increased monotonically with time toward light-curve peak. We estimate an expansion velocity of $2700\pm200$~\kms\ for the photosphere. It is worth mentioning that the derived photosphere expands at a much slower rate than the BALs ($\gtrsim10,000$~\kms) in the UV spectra, which either suggests a difference in the physical location of the outflow or that the two kinematic components are unrelated (see additional discussion in \autoref{sec:photosphere_evo}). If the photosphere was expanding at this constant velocity during the entire rise, this implies that the most bound debris fell back at $t_0 = -30.6$ days, which is well before the most recent ZTF pre-flare upper limit at $\Delta t = -21.7$ days. We estimated a similar reference time of $-26$ days by fitting the luminosity evolution during rise time with a quadratic function $L\propto(t-t_0)^2$. The prediction of a quadratic rise traces back to the ``fireball'' model that is frequently used to describe Type Ia supernova light curves \citep{Riess_1999}. The TDE ASASSN-19bt, which has a densely sampled TESS rising light curve, also exhibits a power-law rise with an exponent of $\sim2$ \citep{2019ApJ...883..111H}.

\subsection{Evolution of the UV Spectra} \label{sec:HST_UV_analysis}
\begin{figure*}
\centering
\includegraphics[width=7in, angle=0]{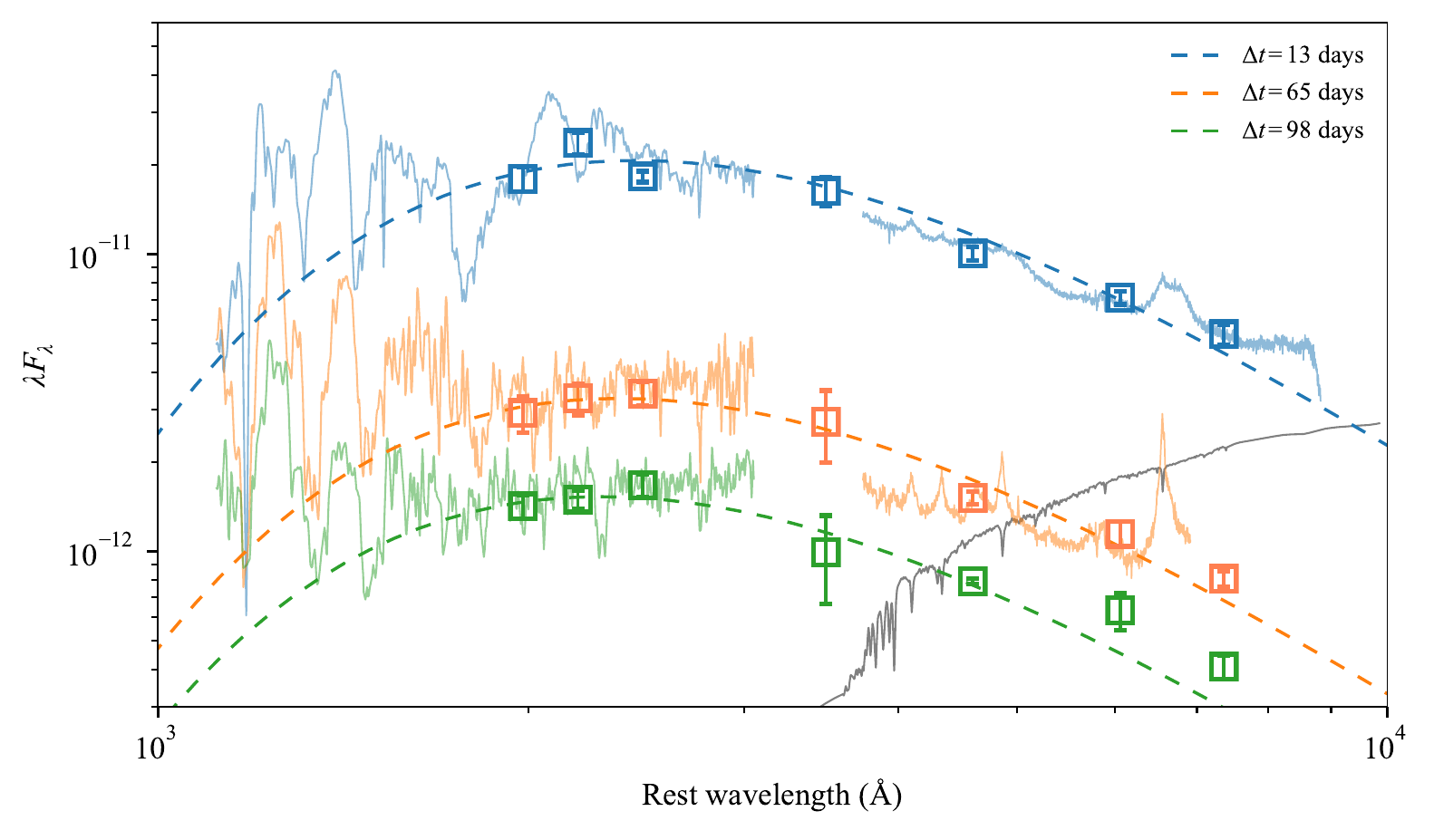}
\caption{A sequence of {\it HST} UV spectra of AT~2019qiz, along with the UV and optical ($UVW2$, $UVM2$, $UVW1$, and $ugri$) photometry (squares) interpolated to the three {\it HST} epochs. Optical spectra obtained on $\Delta t=13$ days (blue) and $\Delta t=71$ days (orange) are also shown. The dashed lines mark the best-fit single-temperature blackbody ($T_{\rm bb} \approx 16,000$\,K) to the interpolated UV and optical photometry at each epoch. The gray line shows the best-fit galaxy model from \texttt{PROSPECTOR}. As demonstrated here, a single-temperature blackbody can often underestimate the TDE continuum in the FUV.
  \label{fig:tde_seds}}
\end{figure*}

A sequence of three {\it HST} UV spectroscopic observations of AT~2019qiz is shown in \autoref{fig:tde_seds}. Among the three {\it HST} epochs, the spectrum from the first epoch exhibits signatures distinct from those in previous TDEs. The apparent reduction in outflow velocity is also seen in the UV spectra for the first time. Below we describe these differences qualitatively and defer the quantitative details to the following subsections.

At all three epochs, high-ionization broad absorption lines (HiBALs) blueward of the rest wavelengths of \NVs, \SiIVs, and \CIVs\ can be readily seen. The HiBAL absorption troughs are blueshifted more in the first {\it HST} epoch ($v\approx15,000$\,\kms\ on $\Delta t=13$ d) and decelerated to $v\approx10,000$\,\kms\ in the later two epochs ($\Delta t=65$~d and 98 d). In addition, the first {\it HST} spectrum is characterized by broad structures in the NUV ($\lambda_{\rm rest}\gtrsim$1650\,\AA). We later identified these broad features to be associated with \AlIII\ and iron (\ion{Fe}{2} and \ion{Fe}{3}) absorption, making AT~2019qiz the first TDE to be detected with Fe and low-ionization broad absorption lines (FeLoBALs). In fact, we find the BAL pattern in AT~2019qiz sharing more similarities to that in a superluminous supernova (SLSN) than in BAL quasars (BALQSOs), which we discuss in more detail in \autoref{sec:UV_BAL_analysis}.

At the first {\it HST} epoch, we also detect narrow absorption lines with a dispersion of $240$~\kms\ at the host redshift $z_{\rm abs}\approx z_{\rm gal}$. The narrow absorption lines are only marginally detected at the later {\it HST} epochs owing to lower S/N. However, for the stronger absorption lines in the FUV, the line strengths do not seem to vary significantly between the first and second {\it HST} epochs.

\subsubsection{Narrow Absorption Lines}
\begin{figure*}
\centering
\includegraphics[width=7in, angle=0]{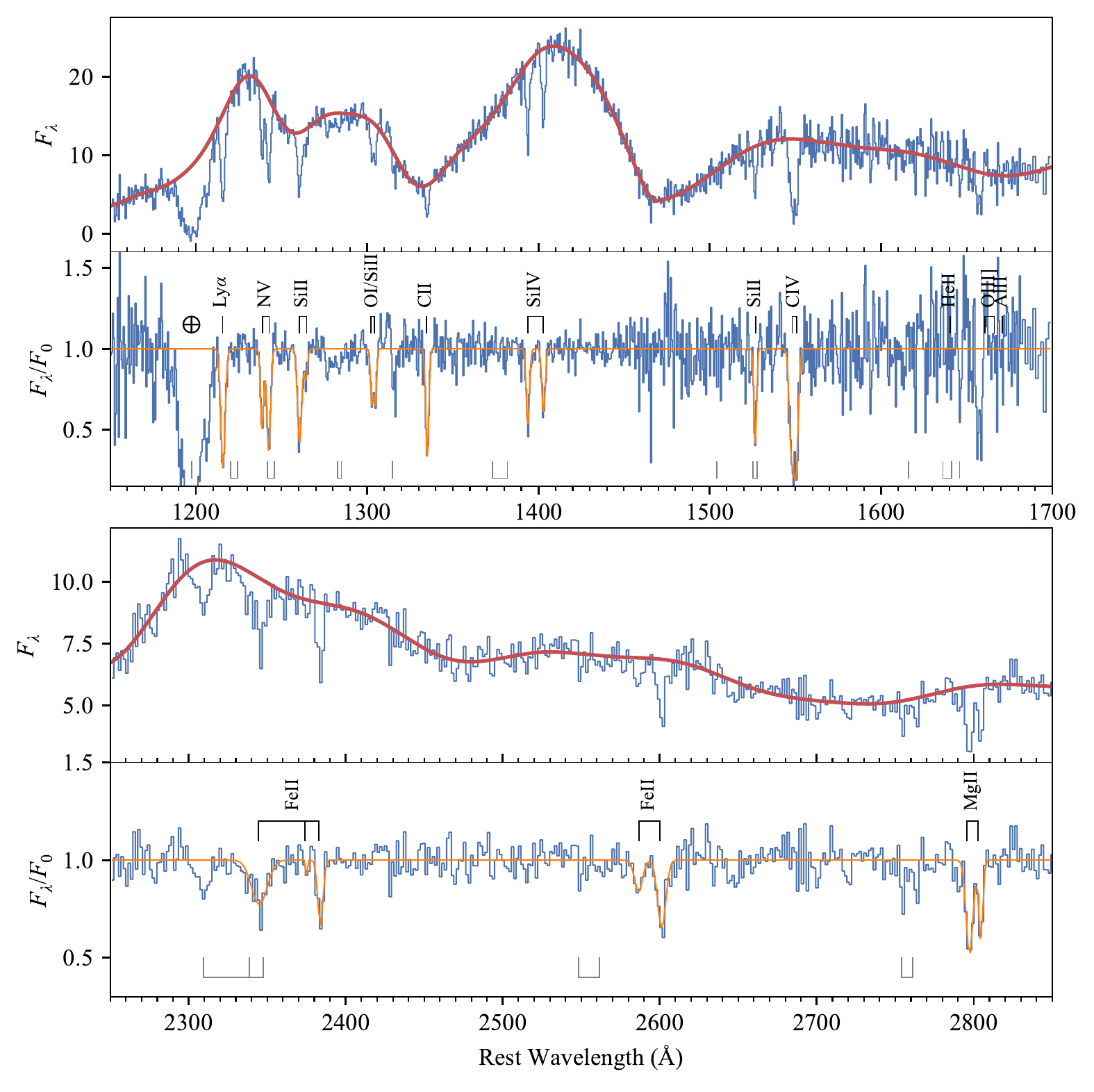}
\caption{Narrow absorption lines in the {\it HST} FUV (top) and NUV (bottom) spectra on $\Delta t=13$ days. For each wavelength segment, the original spectrum (blue) and the smoothed effective continuum (red) are shown in the top panel while the normalized spectrum is shown in the lower panel. We identified the line transitions (labeled in black) and modeled the normalized narrow absorption lines with Gaussian profiles, where the best-fit models are shown in orange.  In the panels with normalized spectra, the gray ticks mark the wavelengths of common Galactic absorption lines.
  \label{fig:uv_narrow}}
\end{figure*}

We identified and measured the narrow absorption of both low- and high-ionization lines in the $\Delta t=13$ d {\it HST} spectrum and tabulated them in \autoref{tab:hst_absorption_lines}. To interpret these narrow lines, we model them by defining the ``effective continuum,'' which consists of the continuum emission and the broad TDE features (emission or absorption), and create a continuum-divided spectrum. This step is done by masking the parts of the spectrum containing narrow absorption lines, then smoothing and interpolating over the masked wavelengths that contain the narrow absorption lines using Gaussian Process with a squared exponential covariance function. We then divided the {\it HST} spectrum by this effective continuum, leaving only the narrow absorption features. Finally, we modeled these absorption lines with Gaussian profiles and measured their line widths and the equivalent widths ($W_r$) in the rest frame. Since the \CIV\ resonance doublet is not resolved in the spectrum, we modeled it by requiring the two components to have the same width. We also fix the width of \ion{O}{1} $\lambda1302$ and \ion{Si}{2} $\lambda1304$ to be the same due to line blending. The smoothed effective continuum, the normalized spectrum, and the best-fit models are shown in \autoref{fig:uv_narrow}. From the line fit, we measured a S/N-weighted velocity offset of $70\pm90$~\kms\ that is consistent with the systemic host velocity using only the isolated and unblended absorption lines. We further place an upper limit of $\lesssim200$~\kms\ on the outflow velocity based on the weak \ion{Si}{2} $\lambda1265$ feature. The average dispersion of the absorption lines is $240\pm70$~\kms.

\begin{figure}
\centering
\includegraphics[width=3.5in, angle=0]{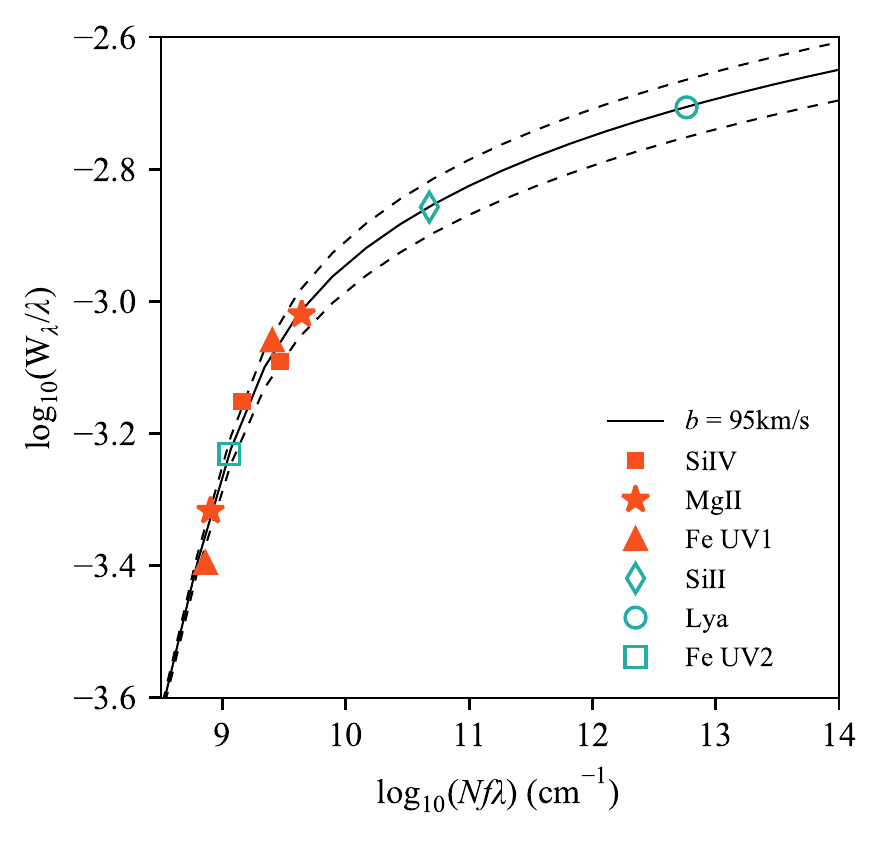}
\caption{The best-fit CoG curve with $b=95$~\kms\ is shown by the black solid curve. The dashed curves mark the $\pm10$~\kms\ uncertainty in $b$. Unblended transitions with multiple components used to constrain the fit are marked by red solid symbols. We interpolate the ionic column density for spectral lines with only a single measured component (green open symbols) to the best-fit curve.
  \label{fig:cog}}
\end{figure}

We adopt the multi-ion single-component curve of growth (CoG) analysis to derive the ionic column densities from the measured $W_r$ of the narrow absorption lines assuming that the Doppler broadening parameter $b$ is the same for each ionic species. This assumption holds if all the absorption, regardless of high- or low-ionization lines, took place in a relatively small region that cannot be resolved by our {\it HST} spectrum.
We solve for $b$ iteratively using the unsaturated, unblended ionic lines with the same lower energy states (\ion{Si}{4}, \ion{Fe}{2}~UV1 , and \ion{Mg}{2}). In our best-fit model (\autoref{fig:cog}), $b = 95$~\kms. The column densities for the uncontaiminated and unblended ionic species are listed in \autoref{tab:hst_absorption_lines}.

We note that one caveat of applying the CoG analysis to a low-resolution spectrum, such as our STIS observation, is that the derived metal column densities can be underestimated by as much as 1 dex \citep{2006ApJ...650..272P}. High-quality and high-resolution spectroscopy of GRB afterglows suggests that the absorption-line structures around strong \ion{Fe}{2} lines are characterized by multiple saturated components spreading over a few hundred \kms. In lower-resolution spectra, these absorption structures may be viewed as a single component with greater effective $b$. Although the larger $b$ value does not necessarily lead to erroneous inference of the column density, it can sometimes affect the measurements by forcing weaker transitions to be optically thin ($\tau_0\propto b^{-1}$) even when they are saturated. Fine structures may also be present in the absorption lines in AT~2019qiz. Therefore, we emphasize that the metal column densities displayed in \autoref{tab:hst_absorption_lines} should be considered as lower limits.

\begin{deluxetable*}{lccccccc}
\tablecaption{Best-fit parameters of narrow UV absorption lines\label{tab:hst_absorption_lines}}
\tablewidth{0pt}
\tablehead{
\colhead{Line} & \colhead{$\lambda_0^a$} & \colhead{$\lambda_{\rm obs}^b$} & \colhead{FWHM} & \colhead{$W_r^c$} & \colhead{$z$} & \colhead{log($gf$)$^d$} & \colhead{log$N$$^e$} \\
\colhead{}     & \colhead{(\AA)} & \colhead{(\AA)} & \colhead{(\AA)} & \colhead{(\AA)} &   &   &\colhead{(cm$^{-2}$)}
}
\startdata
Ly$\alpha$     & 1215.67              & 1215.8    &  3.04$\pm$0.16  &  2.39$\pm$0.15  &  0.0152   &
$-0.08$  & 17.71  \\
\ion{N}{5}     & 1238.82              & 1238.72   &  2.1$\pm$0.21   &  1.13$\pm$0.14  &  0.015    &
$-0.51$  & 14.94     \\
\ion{N}{5}     & 1242.8               & 1242.51   &  2.55$\pm$0.19  &  1.72$\pm$0.15$^f$  &  0.0149   &
$-0.81$  &  \nodata\\
\ion{Si}{2}    & 1260.42              & 1260.55   &  2.79$\pm$0.25  &  1.74$\pm$0.18  &  0.0152   & 0.29   & 15.12    \\
\ion{Si}{2}*    & 1264.73              & 1264.17   &  2.22$\pm$0.66  &  0.45$\pm$0.15  &  0.0146  &    0.55   & \nodata    \\
\ion{O}{1}     & 1302.17              & 1302.72   &  1.72$\pm$0.19  &  0.63$\pm$0.11$^g$  &  0.0155  &   $-0.62$  & \nodata \\
\ion{Si}{2}    & 1304.37              & 1304.72   &  1.72$\pm$0.19  &  0.65$\pm$0.11$^g$  &  0.0154    &
$-0.42$  & \nodata  \\
\ion{C}{2}     & 1334.53, 1335.71     & 1335.07   &  2.12$\pm$0.14  &  1.53$\pm$0.13$^g$  &  0.0155    & $-0.62$   & \nodata  \\
\ion{Si}{4}    & 1393.76              & 1393.83   &  2.23$\pm$0.21  &  1.13$\pm$0.13  &  0.0152   &   0.03   & 14.29 \\
\ion{Si}{4}    & 1402.77              & 1402.94   &  2.3$\pm$0.25   &  0.99$\pm$0.13  &  0.0152   &       $-0.28$  & 14.29   \\
\ion{Si}{2}    & 1526.72              & 1526.53   &  1.77$\pm$0.15  &  1.12$\pm$0.12$^f$  &  0.015   &    $-0.59$  & \nodata \\
\ion{C}{4}     & 1548.20              & 1547.73   &  3.04$\pm$0.18  &  1.96$\pm$0.20$^g$  &  0.0148    & $-0.42$  &  \nodata \\
\ion{C}{4}     & 1550.77              & 1550.42   &  3.04$\pm$0.18  &  2.35$\pm$0.20$^g$  &  0.0149    & $-0.72$  &  \nodata \\
\ion{He}{2}    &   1640.42            &      \nodata &  \nodata &    \nodata &  \nodata \\
\ion{Al}{2}    &   1670.79            &      \nodata &  \nodata &    \nodata &  \nodata \\
\ion{N}{3}]    &   1750.26            &      \nodata &  \nodata &    \nodata &  \nodata \\
\ion{C}{3}]    &   1908.00            &      \nodata &  \nodata &    \nodata &  \nodata \\
\ion{Fe}{2} UV2   & 2344.21              & 2344.99   &  11.77$\pm$0.55 &  2.84$\pm$0.17$^f$  &  0.0154  & $0.04$  & \nodata    \\
\ion{Fe}{2} UV2    & 2374.0               & 2374.99   &  0.68           &  $<$0.06        &  0.0155   &
$-0.55$  & \nodata  \\
\ion{Fe}{2} UV2   & 2382.77              & 2383.86   &  3.9$\pm$0.24   &  1.4$\pm$0.11  &  0.0156    &    0.59   & 13.09  \\
\ion{Fe}{2} UV1   & 2586.65              & 2586.47   &  5.95$\pm$0.61  &  1.04$\pm$0.13  &  0.015    &
$-0.19$  & 13.63  \\
\ion{Fe}{2} UV1   & 2600.17              & 2601.24   &  6.06$\pm$0.28  &  2.27$\pm$0.13  &  0.0155   &    0.35   & 13.63 \\
\ion{Mg}{2}   & 2795.53              & 2797.34   &  5.13$\pm$0.19  &  2.67$\pm$0.12  &  0.0158    &   0.53  & 13.66\\
\ion{Mg}{2}    & 2802.7               & 2804.44   &  3.0$\pm$0.17   &  1.35$\pm$0.09  &  0.0157    &
$-0.21$ & 13.66\\
\enddata
\tablecomments{From the $\Delta t=13$ days {\it HST} spectrum. Absorption lines are modeled as individual Gaussian profiles after division by the local continuum.
$^a$The rest wavelength of the line transition. $^b$The observed line center. $^c$The equivalent width of the line. $^d$Oscillator strength from the atomic spectral line database \citep{1995KurCD..23.....K}.
$^e$Column density.
$^f$The transition may be subject to contamination by foreground absorption. $^g$Blended with neighboring lines in the same absorbing system. Therefore, we do not use the measured $W_r$ to infer the metal column density.}
\end{deluxetable*}

\subsubsection{Identification of the UV Broad Lines} \label{sec:UV_BAL_analysis}
Before delving into the identification of broad features in AT~2019qiz, we remind the readers that the continuum placement, especially in the FUV, may be quite uncertain. Traditionally, we find that the NUV and optical photometry can be described well by a single-temperature blackbody with $T_{\rm bb}\approx$~a few~$\times10^4$~K. This has been confirmed to be generally true by past {\it HST} observations, where the NUV spectra of most TDEs are characterized by a featureless continuum with a slope that is consistent with the \swift\ photometry in each event \citep{Cenko2016,2018MNRAS.473.1130B,2019ApJ...873...92B,2019ApJ...879..119H}. At $\lambda_{\rm rest} < 1600$\,\AA, the common presence of HiBALs often eclipses away a large amount of flux in the FUV continuum and broad emission lines in TDEs, making the FUV continuum placement especially difficult. It is noticed that the FUV continuum can be significantly underestimated by extrapolating the NUV blackbody spectrum, as was found in AT~2018zr \citep{2019ApJ...879..119H} and also in AT~2019qiz (\autoref{fig:tde_seds}). Therefore, in the analysis we avoid measuring line properties that are dependent on the continuum, such as the line flux or balnicity index (BI; a measure of the strength of the broad absorption line features defined in \citealt{1991ApJ...373...23W}) except for the narrow absorption lines, where the effective continuum can be determined relatively accurately.

\begin{figure*}
\centering
\includegraphics[width=7in, angle=0]{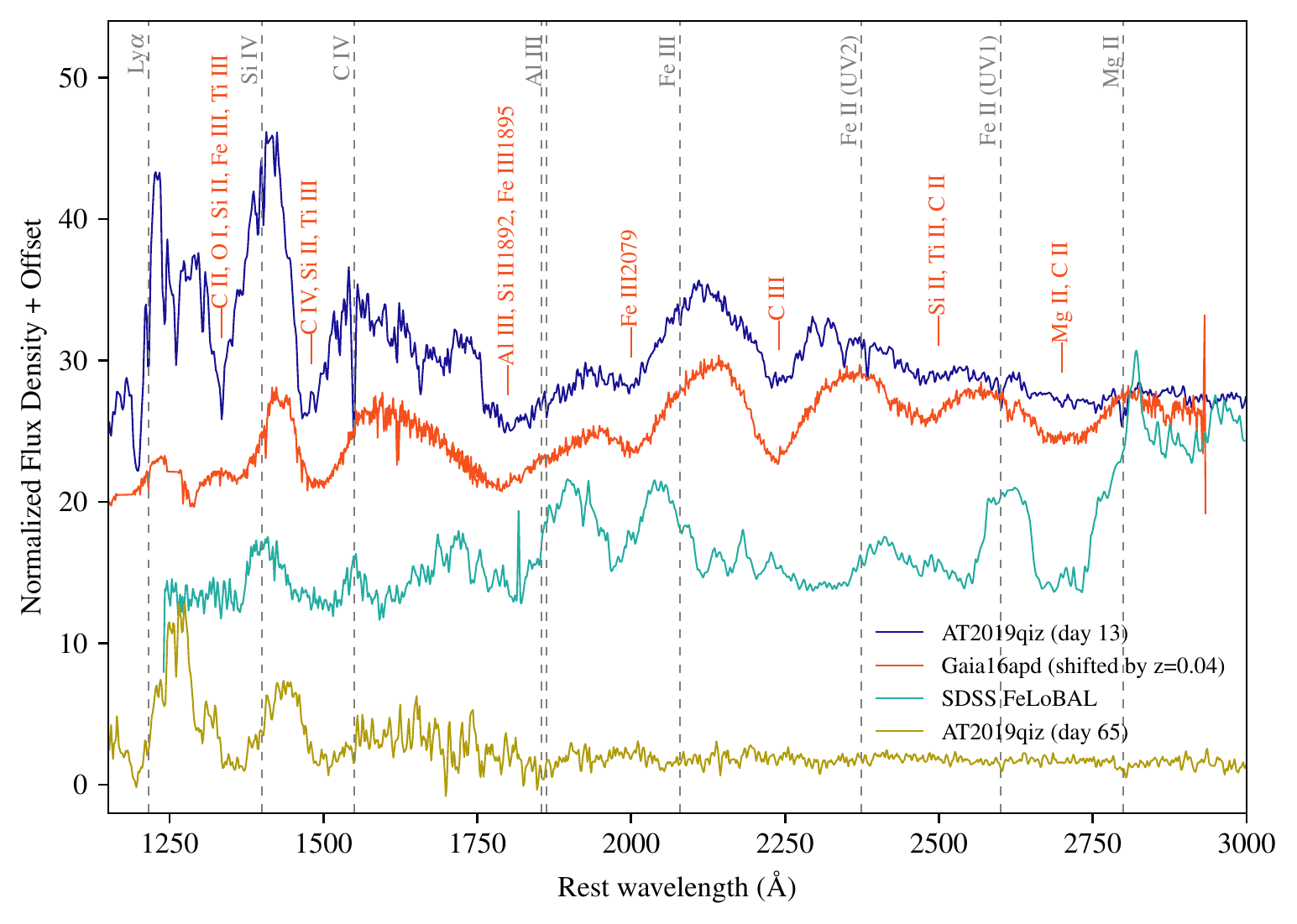}
\caption{UV spectra of AT~2019qiz on day 13 and day 65, compared to the SLSN Gaia16apd \citep{2017ApJ...840...57Y}, and the FeLoBAL QSO SDSS~J090152.04+624342.6 \citep{2017FrASS...4...40W}. The first UV spectrum of AT~2019qiz bears a stronger resemblance to that of Gaia16apd. The orange labels mark the potential contributing species identified by \cite{2017ApJ...840...57Y}.
  \label{fig:lineid}}
\end{figure*}

Despite the fact that the blackbody spectrum extrapolated from NUV and optical observations is clearly underestimating the FUV continuum, we keep them in \autoref{fig:tde_seds} to guide the eye.
Broad emission and absorption features are detected at all three epochs. In order to identify the lines, it is natural to compare the {\it HST} spectra of AT~2019qiz with a BALQSO since both TDEs and BALQSOs are phenomena driven by accretion and winds. Furthermore, similarities between the rest-frame UV spectra of TDEs and BALQSOs  have been drawn in the past \citep[e.g.,][]{2019ApJ...873...92B}. We find broad structures in both the FUV and NUV sides of the spectrum on day 13. Broad emission lines seem to have formed around the rest wavelengths of the high-ionization lines \NVs, \SiIVs, and \CIVs. Blueward of these emission lines are the absorption troughs associated with these high-ionization transitions. We estimate the blueshift velocity of the broad absorption features ($v_{w}$) with respect to the minimum of the troughs. The broad features associated with \SiIVs\ and \CIVs\ are blueshifted by $v_{w} \approx 15,000$ \kms, whereas the velocity offset of the \NVs\ absorption trough cannot be precisely measured owing to overlap with the geocoronal emission.

A subclass of BALQSOs, the low-ionization broad absorption line (LoBAL) QSOs, have UV spectra imprinted by broad absorption of \MgII, \ion{Al}{3}, and \ion{Al}{2} at $\lambda_{\rm rest} > 1750$\,\AA. In even rarer cases, iron absorption lines are present in the QSO spectra, and thus these QSOs are termed ``FeLoBAL'' QSOs. Given the presence of similar ionic species, we compare AT~2019qiz with the spectrum of a SDSS FeLoBAL QSO at $z \approx 2.1$ in \autoref{fig:lineid}. From this comparison, we infer that the absorption around 1800\,\AA\ in AT~2019qiz is likely due to a blueshifted ($v_{w} \approx 9000$ \kms) \AlIII\ line. At $\lambda_{\rm rest}>2000$\,\AA, the $\Delta t=13$ d spectrum of AT~2019qiz is characterized by several broad peaks and valleys around the most prominent \ion{Fe}{2} and \ion{Fe}{3} lines. The iron lines that are possibly associated with the broad NUV features are labeled by dashed lines in \autoref{fig:lineid}. If we attribute the absorption features at 2000\,\AA\ and 2240\,\AA\ to \ion{Fe}{3} and \ion{Fe}{2} UV2, the measured wavelengths correspond to $v_{w} \approx 12,000$--16,000\,\kms, which is in good agreement with that derived from the HiBALs. Unlike the BALQSOs, the {\it HST} spectra of AT~2019qiz did not show any significant broad, blueshifted \MgII\ absorption at any phase (\autoref{fig:lineid}).

Intriguingly, we find the $\Delta t=13$ days UV spectrum of AT~2019qiz to bear greater resemblance to the SLSN Gaia16apd \citep{2017ApJ...840...57Y} than to any type of BALQSOs (\autoref{fig:lineid}). The similarity between the two spectra  possibly stems from the fact that both events had a similar blackbody temperature ($T_{\rm bb} \approx 17,000$\,K) and an expanding photosphere, as probed by the SED fit in \autoref{sec:sed_fit}, during the early phases. The absorption troughs at $1750<\lambda_{\rm rest}<2400$\,\AA\ aligned particularly well with Gaia16apd after redshifting the SLSN spectrum by 12,000\,\kms. Using the \texttt{syn++} model that generates synthetic spectra for homologously expanding atmospheres, \cite{2017ApJ...840...57Y} identified the ionic species that contribute to each absorption line in Gaia16apd. This new comparison makes a difference in the identification of the feature at 1800\,\AA\ and 2240\,\AA. If the ionization state is also similar in AT~2019qiz and Gaia16apd, the 2240\,\AA\ feature would instead be attributed to the absorption of \ion{C}{2} with $v_{w} \approx 11,000$\,\kms. In addition, the 1800\,\AA\ feature would not be uniquely associated with \ion{Al}{3}, but also with \ion{Si}{2} and \ion{Ti}{2}. The absorption troughs around the \SiIVs\ and \CIVs\ lines may also be contributed by other low-ionization species such as \ion{C}{2} and \ion{Si}{2}. While many LoBALs are detected in the day 13 UV spectrum, the lack of \MgII\ in AT~2019qiz would possibly need to be explained by the composition.

\begin{figure*}
\centering
\includegraphics[width=7in, angle=0]{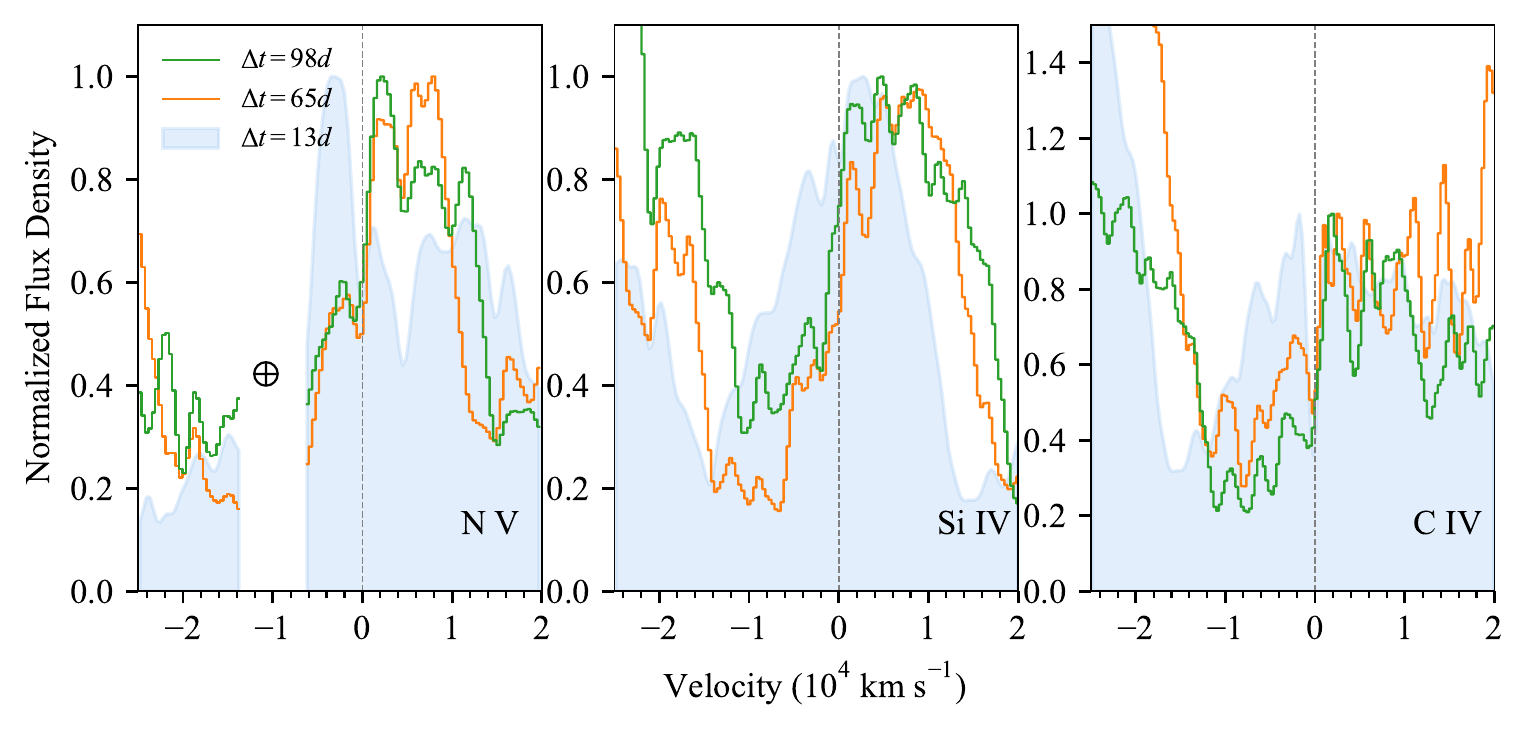}
\caption{Three epochs of {\it HST} STIS spectra in velocity space with respect to the rest wavelengths of the high-ionization lines \ion{N}{5}, \ion{Si}{4}, and \ion{C}{4}. The region affected by geocoronal airglow emission indicated by a circled plus symbol is truncated. The UV spectrum from the first {\it HST} epoch ($\Delta t=13$\,d) is shaded in pale blue while the spectra from the second and the third epochs are plotted as orange and green curves, respectively. It can be seen that the absorption troughs shifted to lower velocities as time progressed from 13\,d to 65\,d, which makes the emission lines seem more redshifted.
  \label{fig:UV_lines_vel}}
\end{figure*}

The broad UV features evolved significantly from day 13 to day 65. The most remarkable difference is that the absorption troughs associated with the high-ionization \SiIVs\ and \CIV\ lines became redder, decelerating from $v_w\approx15,000$ \kms\ to $v_w\approx10,000$ \kms\ (\autoref{fig:UV_lines_vel}). We also notice that the NUV spectrum became featureless and the UV spectra resemble those of previously observed TDEs more at later epochs. The spectra of day 65 and day 98 are very similar except that the bluest \SiIVs\ absorption edge seems to have moved to a slightly lower velocity on day 98 (\autoref{fig:UV_lines_vel}).

\subsection{Optical Spectroscopic Analysis}
\begin{figure*}
\centering
\includegraphics[width=6in, angle=0]{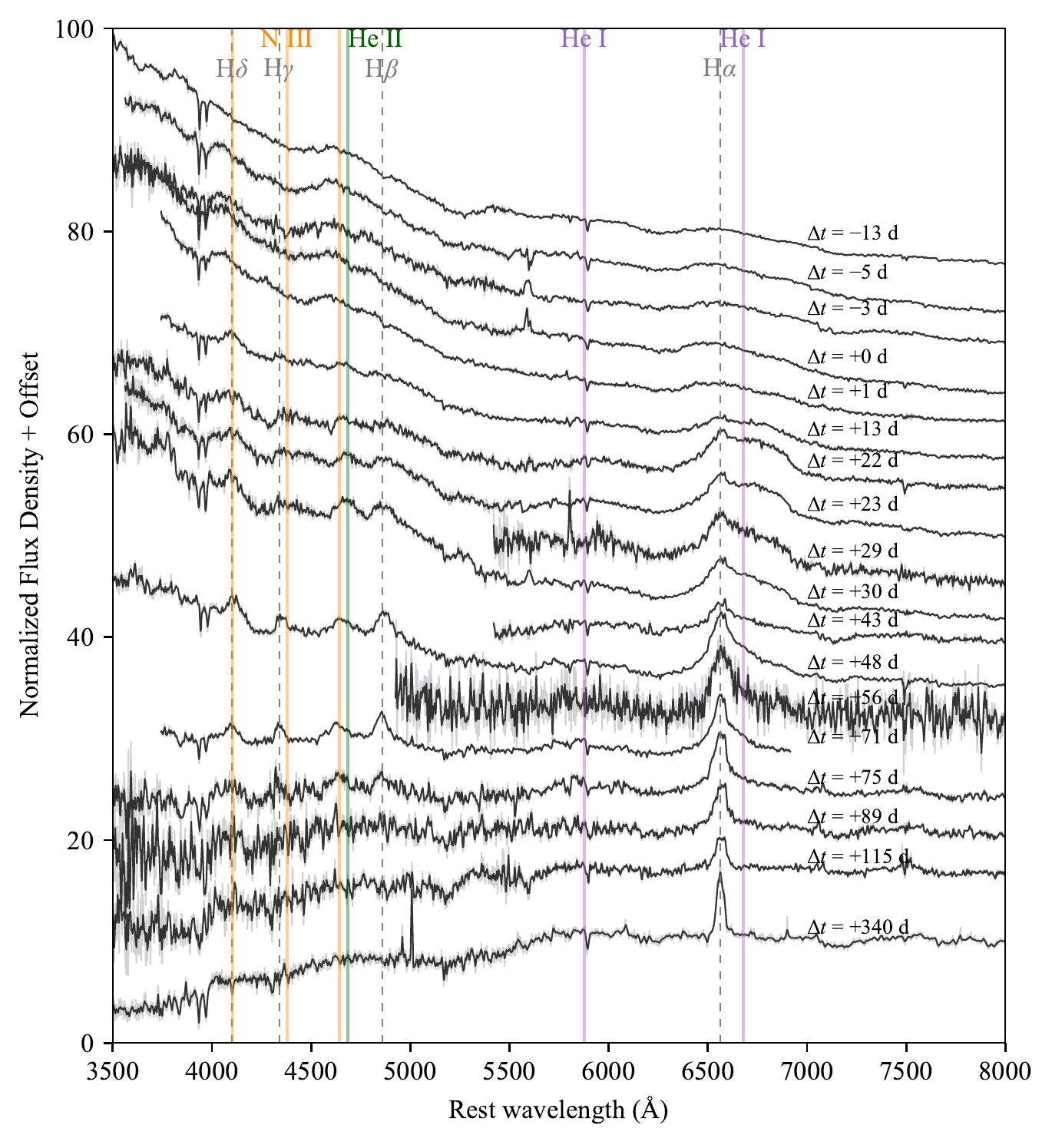}
\caption{Optical spectra of AT~2019qiz obtained between $\Delta t=-13$ days and $\Delta t=340$ days. The spectra smoothed by a Gaussian kernel with $\sigma=2$~\AA\ are shown in black lines while the original data are shown in gray. The gray dashed lines mark the wavelengths of the Balmer series. Colored vertical lines mark the wavelengths of \ion{N}{3} (orange), \ion{He}{2} (green), and \ion{He}{1} (purple). Significant line-profile evolution is seen in the \Ha\ emission.
  \label{fig:optspec_all}}
\end{figure*}

\begin{figure}
\centering
\includegraphics[width=3.5in, angle=0]{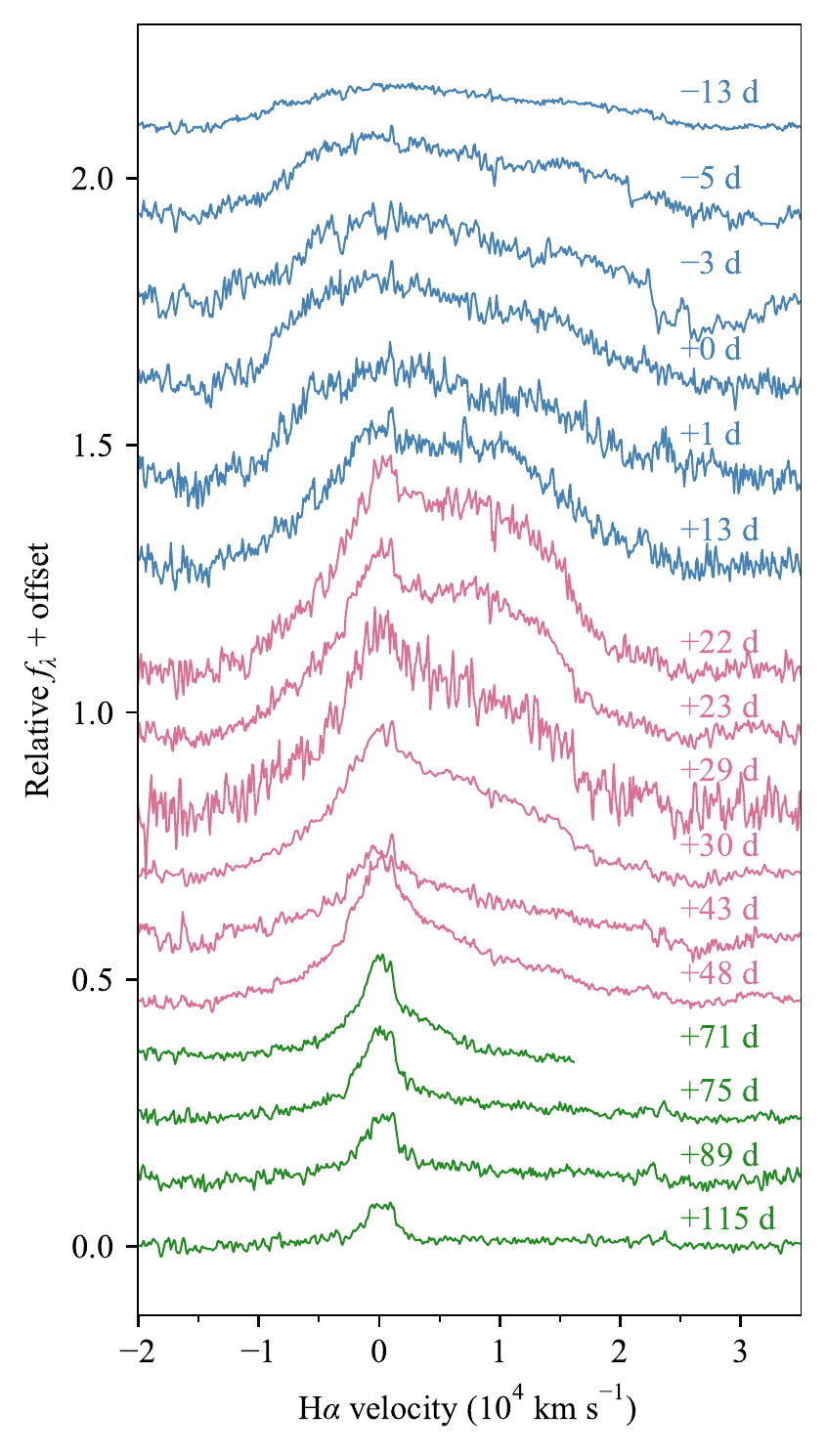}
\caption{Evolution of the \Ha\ emission profile after continuum subtraction. We used models with a different number of Gaussian components to fit the line at different phases. The \Ha\ line profile in the earliest spectra can be characterized by two broad Gaussians (blue). In the post-peak period, a narrow component became apparent, which motivated us to adopt a three-component model (pink) to describe the line shape. In later spectra, the red shoulder in the broad \Ha\ disappeared; thus, we use a narrow and a broad component centered around \Ha\ to model the emission line (green).
  \label{fig:h_alpha_evo}}
\end{figure}

We show a sequence of 18 optical spectra of AT~2019qiz in \autoref{fig:optspec_all}.
The spectroscopic features in AT~2019qiz underwent significant changes on a timescale of days to weeks. This is best illustrated by the evolution in the shape of the \Ha\ emission (\autoref{fig:h_alpha_evo}). The pre-peak and early post-peak spectra from day $-13$ to day 13 are characterized by a blue continuum with very broad emission features around \Ha\ and \HeII. The latter is likely contaminated by the emission from \Hb\ as the entire broad feature spans rest wavelengths 4400--5200\,\AA. During this stage, the \Ha\ emission structure has a full width at half-maximum intensity (FWHM) of $2.6\times10^{4}$~\kms. We fit this broad \Ha\ feature with a double Gaussian function and derived two components with similar FWHM of $\sim1.5\times10^{4}$ \kms\ centered roughly at the rest wavelength of \Ha\ and at a redshift of $1.6\times10^{4}$ \kms. The red component has a weaker line flux about half that of the blue component.

On day 22, the \Ha\ emission-line profile changed dramatically, with a narrow component appearing around the rest-frame \Ha\ wavelength. We then modeled the \Ha\ emission with three Gaussians simultaneously while requiring the narrow component to have FWHM $<6,000$~\kms. The best-fit model consists of a broad \Ha\ base, which is the sum of two Gaussians as in the earlier epochs, with a FWHM of $2.1\times10^4$~\kms and a narrow \Ha\ component with a FWHM of $3,000$~\kms\ centered at zero-velocity. During this time, the blue and red component of the broad \Ha\ emission became almost equally strong while the velocity offset of the red component decreased to $1.1\times10^4$~\kms.

The broad \Ha\ base continues to decay while the narrow \Ha\ remains strong in all later epochs. The broad \Ha\ has not only showed a decrease in line flux but also a change in the line shape. From day 71 onward, the broad \Ha\ shifted to a single Gaussian centered around the rest \Ha\ wavelength where the red shoulder detected with $v>10,000$\,\kms\ at previous epochs disappeared. The FWHM of the broad \Ha\ dropped to $\sim12,000$\,\kms. The FWHM of the narrow \Ha\ emission fluctuates between 3000 and 4000\,\kms\ since it first appeared around day 22.

Narrow Balmer emission lines from higher excited states and narrow Bowen emission (\ion{He}{2}+\ion{N}{3}) likely also appeared from day 22 onward.  However, only after day 49 did the broad emission and continuum fade enough for the narrow lines to be detected. The presence of both the Balmer lines and \ion{He}{2}+\ion{N}{3} makes AT~2019qiz a TDE-Bowen, which is a spectral class that makes up about half of the optical TDE population defined by \cite{2020arXiv200101409V}. On day 71 when the narrow emission lines are the strongest, we measured a FWHM of 4000\,\kms\ for \Hb\ and a FWHM of 3100\,\kms\ for H$\gamma$. The line intensity ratio \Ha/\Hb\ $\approx 1.2$ corresponds to a flat Balmer decrement, which has also been reported in several TDEs \citep[e.g., AT~2018hyz;][]{2020apJ...903...31H,2020arXiv200305470S}. We also measured intensity ratios H$\gamma$/\Hb\ $\approx 0.6$ and (\ion{He}{2}+\ion{N}{3})/\Hb\ $\approx1$.

\subsection{Host Galaxy and the X-ray Emission} \label{sec:host}
The host galaxy of AT~2019qiz is 2MASX $J04463790-1013349$ (WISEA J$044637.88-101334.9$) at $z=0.0151$. The $r$-band Pan-STARRS image reveals that it is a spiral galaxy with a clear bar structure. \citetalias{2020mnRas.499..482N} measured a Sersic index in the range 5.2--6.3, which is considered high compared to the galaxy sample in the same blackhole mass bin \citep{2017ApJ...850...22L}. Like many other TDEs, AT~2019qiz is in a galaxy with a more concentrated population of stars \citep{2020ApJ...891...93F}.

\begin{figure*}
\centering
\includegraphics[width=7in, angle=0]{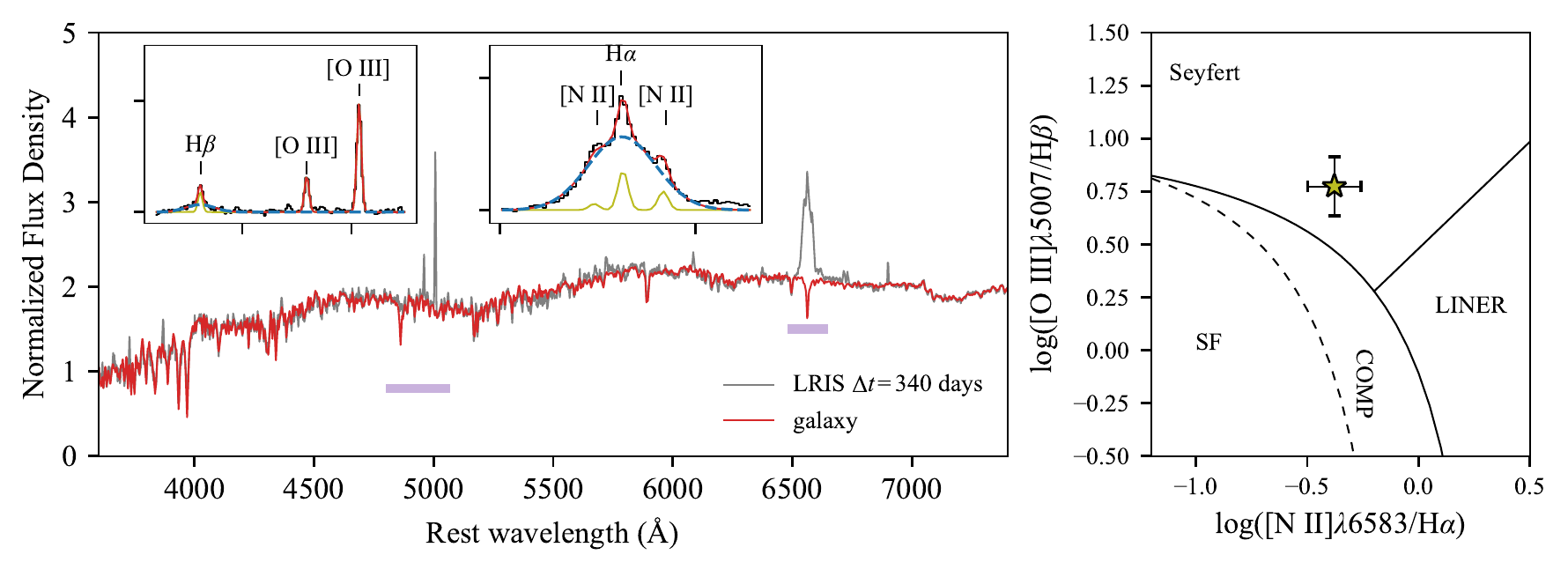}
\caption{{\it Left:} Late-time Keck spectrum of AT~2019qiz (gray) overlaid with the best-fit host-galaxy model (red) with pPXF. We fit the stellar population as well as the host and TDE emission lines simultaneously. The emission-line spectrum (host-subtracted) around \Ha\ and \Hb\ are shown in the insets, where the TDE emission is marked by blue dashed lines and the host emission is plotted in yellow. {\it Right:} The BPT emission-line diagnostic diagram. The narrow-line ratios of the host galaxy of AT~2019qiz fall in the region occupied by AGNs.
  \label{fig:host_spec_bpt}}
\end{figure*}

In the late-time optical spectra of AT~2019qiz, we detect nebular emission lines [\ion{O}{3}] $\lambda\lambda4959, 5007$ and [\ion{N}{2}] $\lambda\lambda$6548, 6583 that are indicative of pre-flare active galactic nucleus (AGN) activity. Given that the TDE light has faded substantially in the $\Delta t=340$ days spectrum, we fit its stellar and gas kinematics simultaneously using the Penalized PiXel-Fitting (pPXF) package \citep{2017MNRAS.466..798C} with the MILES stellar library (\autoref{fig:host_spec_bpt}). In addition to the narrow nebular emission lines, we added a second gas component in our fit to account for the TDE Balmer emission at late times, which has also been observed in previous TDEs \citep[e.g., AT2018zr;][]{2019ApJ...879..119H}. Our best-fit width for the TDE Balmer lines (FWHM $\approx 2000$~\kms) is consistent with the width of the narrow \Ha\ component in earlier TDE spectra ($22\leq\Delta t \leq115$ days). We measured the flux of the host-galaxy emission lines and plot the line ratios in a Baldwin, Phillips, \& Terlevich (BPT; \citealt{1981PASP...93....5B}) diagram (\autoref{fig:host_spec_bpt}). The line ratios are in good agreement with the nebular emission lines being produced by AGN photoionization.

During the flare, \citetalias{2020mnRas.499..482N} measured a weak X-ray luminosity of $L_X=5.1\times10^{40}$~erg~s$^{-1}$ in AT~2019qiz that is 2--3 orders of magnitude below the UV and optical luminosity. \citetalias{2020mnRas.499..482N} associated the X-ray emission with the TDE rather than the AGN owing to variability in the X-ray flux and the evolution of the hardness ratio. However, the hardness ratio of $-0.1\pm0.04$ from the merged XRT observations is on the higher end of the TDE distribution when compared to X-ray bright TDEs such as ASASSN-14li, and is more similar to those seen in AGNs \citep{2018ApJ...852...37A}. Therefore, the possibility of AGN-driven X-ray emission in AT~2019qiz cannot be completely ruled out.

\section{Discussion} \label{sec:discussion}
\subsection{Photosphere Evolution}\label{sec:photosphere_evo}
The UV and optical photometry reveals the evolution of the blackbody radius from the pre-peak phase out to $\Delta t \approx 100$ days. The blackbody radius was initially expanding linearly with a velocity of 2,700\,\kms\ up to the time of peak light in the UV and optical. The constant-velocity phase was then followed by a constant-radius phase for $\sim25$ days before the radius began to shrink with time like that observed in previous TDEs, as a result of the combination of decreasing luminosity and constant color temperature.

Our intensive follow-up observations of the optical photosphere and the BALs in the UV spectra strongly support the presence of an outflow that evolved significantly with time in AT~2019qiz. Various models have been proposed to explain the origin of TDE outflows \citep{2009MNRAS.400.2070S,2011MNRAS.410..359L,2018ApJ...859L..20D,2020MNRAS.492..686L}.
In such models, the wind typically has a nonspherical configuration, and the velocity can have some variation depending on the wind-launching radius and observer inclination. However, when strong optical emission is observed, one can assume that the wind is mass-loaded and optically thick along the line of sight, thus justifying the adoption of a quasispherical model for studying the wind physics \citep{2016MNRAS.461..948M,Roth2016}. Here we construct a simple 1D wind model, where the mass outflow rate $\dot{M}_{\rm wind} = 4\pi r^2 \rho_{\rm wind} v_{\rm wind}$, based on the observed results.
The photospheric radius ($R_{\rm ph}$) corresponds to the surface of electron scattering where the optical depth $\tau_e\propto \rho(r=R_{\rm ph})\,R_{\rm ph}=2/3$. Therefore, $R_{\rm ph}$ should be proportional to the mass outflow rate divided by the wind velocity, $R_{\rm ph}\propto \dot{M}_{\rm wind} / v_{\rm wind}$.
If we assume that the observed luminosity scales roughly linearly with the accretion rate $\dot{M}_{\rm acc}$, then from the three UV BAL velocity measurement at $\Delta t$= 13, 65, and 98 days post-peak, we can derive a scaling relation $v_{\rm wind} \propto \dot{M}_{\rm acc}^{0.2}$. Putting this together with the observed photosphere radius evolution $R_{\rm bb} \propto \dot{M}_{\rm acc}^{0.6}$ during these times, we can constrain that $\dot{M}_{\rm wind} \propto \dot{M}^{0.8}_{acc}$. This is consistent with the results of \cite{2009MNRAS.400.2070S} and \citet{2011MNRAS.410..359L}, stating that a stronger wind is launched at earlier phases in TDEs when the mass fallback rate is higher.

In the analytical model developed by \cite{2016MNRAS.461..948M}, only a small fraction of the bound debris accretes onto the black hole. The binding energy of the accreted mass naturally gives rise to a quasispherical outflow that is massive enough to absorb and reprocess the hard EUV/X-ray photons into UV and optical emission.
One prediction of the \cite{2016MNRAS.461..948M} model is that the TDE light curve can deviate from the mass fallback rate ($\propto t^{-5/3}$) at early times owing to photon trapping. They find that the observed luminosity can be suppressed at first because of adiabatic losses in the inner wind, resulting in a flatter light curve than $\propto t^{-5/3}$. The suppression continues until the trapped photons are advected to a radius where the photon diffusion time is short compared to the outflow expansion time on a timescale of $t_{\rm tr}/t_{\rm fallback}\approx2.6\beta^{3/5}M_{\rm bh}^{-1/2}m_\star^{1/5}R_{\rm in,6}^{-3/5}$. At this point, photons can diffuse out of the ejecta freely and thus allow the photosphere to cool. The effect of photon trapping aligns with the observed $T_{\rm bb}$ evolution of AT~2019qiz, where the measured temperature started to decrease at around one fallback time. This may also explain why the cooling in ASASSN-14ae and ASASSN-19bt occurred at different phases, since the trapping time is dependent on other parameters.

The balance between a decreasing luminosity and temperature cooling gives rise to the plateau in the blackbody radius from $\Delta t = 0$ to 25 days. In practice, the photosphere velocity does not need to follow the outflow velocity if the density or ionization of the photosphere are changing. That is, even if the gas producing the UV and optical continuum continues expanding at a constant velocity, a reduction in the electron density can move the electron-scattering photosphere inward, and leads to a plateau or even a decrease in the derived blackbody radius as observed in AT~2019qiz.

\subsection{TDE UV Spectra}
Blueshifted broad absorption features detected in AT~2019qiz signify the presence of outflowing gas at a phase as early as $\Delta t=13$ days. The ejecta producing the BALs would have reached a distance of $\sim5\times10^{15}$~cm if they have been moving at the same velocity since the most bound debris returned to the pericenter (assuming $t_\mathrm{fallback} = -25$ to $-30$ days). BALs have also been detected at different stages in almost all of the TDEs observed with UV spectroscopy, including PS1-11af, iPTF15af, iPTF16fnl, and AT2018zr \citep{Chornock2014,2019ApJ...873...92B,2018MNRAS.473.1130B,2019ApJ...879..119H}. Among the {\it HST} TDE sample, ASASSN-14li is the only object that exhibits pure broad emission lines \citep{Cenko2016}.

To this date, iPTF16fnl ($\Delta t=7, 22, 44$ days), AT~2018zr ($\Delta t=23$, 36, 41, 59, 62 days), and AT~2019qiz ($\Delta t=13$, 65, 98 days) are the only TDEs with a sequence of {\it HST} UV spectra \citep{2018MNRAS.473.1130B,2019ApJ...879..119H}. Among these, AT~2019qiz has the longest observational baseline. Each of these TDEs undergoes a very different evolution path. In the earliest {\it HST} epoch, iPTF16fnl was observed with weak HiBALs (FWHM $\approx 6,000$\,\kms) that diminished over time. AT2018zr started resembling ASASSN-14li with a UV spectrum characterized by broad emission lines. Broad absorption troughs blueward of both high- and low-ionization species with a velocity offset of $\sim0.05c$ started to appear from day 59 onward. AT~2019qiz exhibits a rare FeLoBAL spectrum at the first epoch, which then transitioned into a HiBAL spectrum at later two epochs. The HiBALs remained strong in AT~2019qiz though the outflow velocity decreased from 15,000 to 10,000\,\kms\ in $\sim50$ days (\autoref{fig:UV_lines_vel}). The decrease in outflow velocity is consistent with the theoretical prediction, where the wind becomes weaker with time as the mass fallback/accretion rate declines.

\begin{figure*}
\centering
\includegraphics[width=7in, angle=0]{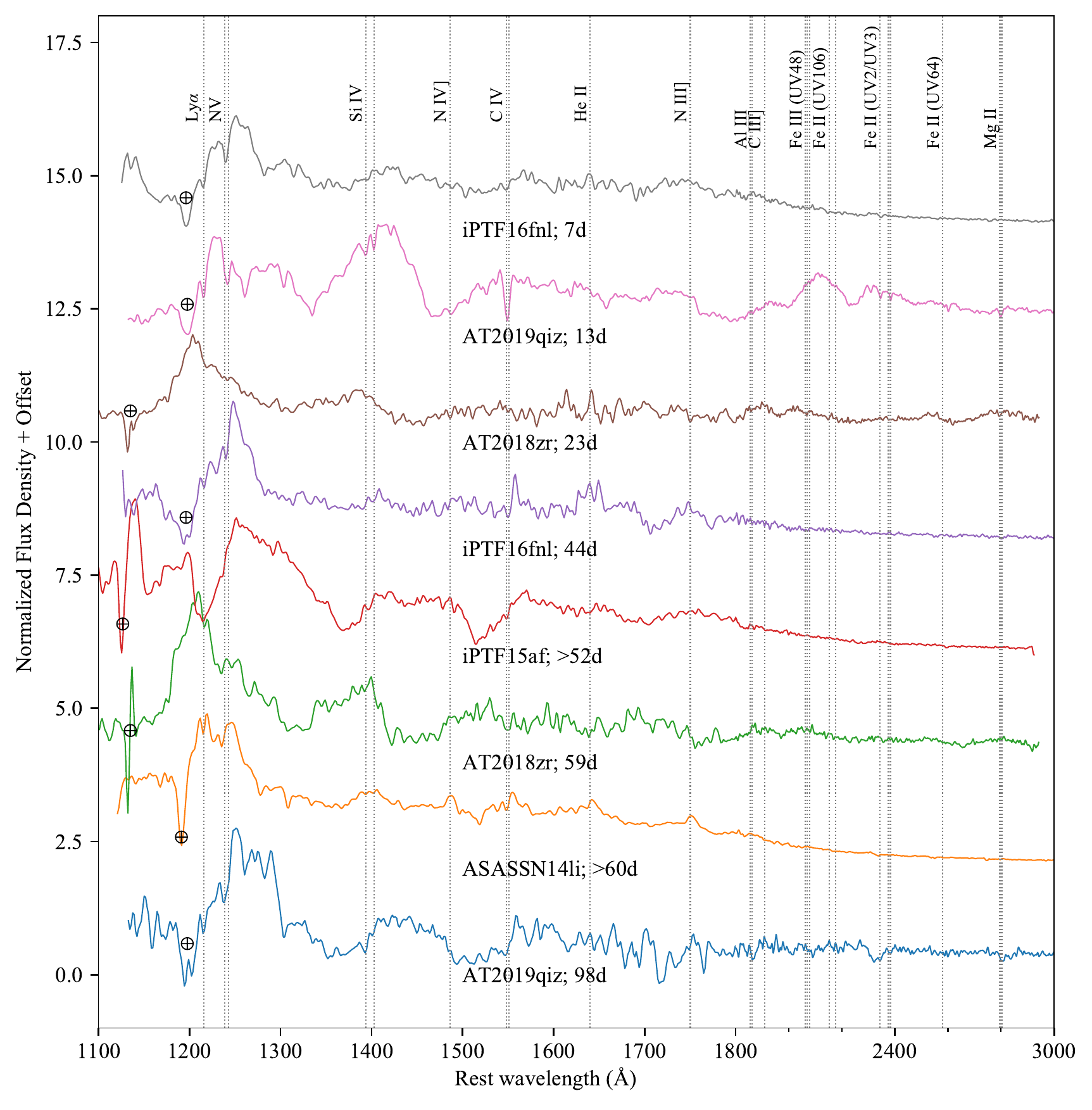}
\caption{A compilation of TDE UV spectra normalized to rest wavelength 1700\,\AA. Most of the TDE UV spectra show fast-moving ($v\approx10^3$--$10^4$~\kms) BAL features that signify the presence of outflows at some point of their evolution. The BAL features in TDEs are typically seen around high-ionization lines such as \CIV\ and \SiIV\ yet are usually absent in their NUV spectra. The $\Delta t=13$ days spectrum of AT2019qiz is distinct from other known TDEs by having broad Fe features in the NUV. Note that the scale on the abscissa has been compressed for the NUV segment (1800--3000\,\AA) to allow more detailed examination of the FUV portion of the spectra.
  \label{fig:uv_spec_compare}}
\end{figure*}

We select the most representative phases from the multi-epoch {\it HST} spectra of iPTF16fnl, AT2018zr, and AT2019~qiz and compare them with the single-epoch spectra obtained for all the rest of the TDEs shown in \autoref{fig:uv_spec_compare}. While we arrange the spectra in \autoref{fig:uv_spec_compare} with respect to phase, we find that any two spectra obtained at a comparable phase do not guarantee similarity.

The spectrum that stands out in the current UV TDE population is the AT~2019qiz UV spectrum at $\Delta t=13$ days (\autoref{fig:uv_spec_compare}). The presence of FeLoBALs in the UV spectrum of AT~2019qiz is in accordance with the reprocessing scenario. Photoionization models of FeLoBAL QSOs, which are typically X-ray faint, suggest that high column densities are required in order to produce the iron features \citep[e.g., $N_H \approx 10^{20.6}$~cm$^{-2}$;][]{2008ApJ...688..108K}. Analogously, the BAL clouds need to be shielded from the hard X-ray photons to form the FeLoBAL in AT~2019qiz. The shielding gas may be an inner wind consisting of material that has fallen back more recently as proposed in the \cite{2016MNRAS.461..948M} model. In a few fallback times, the column density drops sufficiently low such that the ejecta become transparent to the X-ray and EUV photons. This likely explains why the iron and low-ionization features disappeared, leaving only the HiBALs in later {\it HST} spectra of AT~2019qiz.

As shown in \autoref{fig:lineid}, this spectrum is more like that observed in SLSNe instead of in TDEs or BALQSOs. According to the \cite{2016MNRAS.461..948M} model, the outflow properties of TDEs may be similar to those in engine-driven supernovae as black hole accretion continues to inject energy into the ejecta. The observed similarities between the early-time UV spectral features in AT~2019qiz and that in the SLSN Gaia16apd are strong evidence supporting this connection.

The HiBAL TDE spectra, which are seen in at least half of the TDEs at a later phase, all seem to be very similar except for small differences in outflow velocities (6,000--15,000\,\kms). The reduction in the mass fallback rate and the expanding ejecta likely facilitated changes in the ionization structure and the column densities that cause the UV spectral features to transition from SLSN-like to more BALQSO-like. It is worth noting that the \CIVs\ and \SiIVs\ emission always appear to be weaker than the \NVs\ emission in TDEs than that in AGNs and BALQSOs. This may be attributed to the stellar debris having a higher N/C ratio owing to CNO processing in the stellar core \citep{2016MNRAS.458..127K,Cenko2016,2017ApJ...846..150Y,2018ApJ...857..109G,2019ApJ...882L..25L,2020arXiv200710996L}.

The diverse properties of the TDE UV spectra perhaps reflect the intricate processes involved in a TDE such as accretion disk formation and wind launching. In recent theoretical developments, studies have shown that orientation effects can, at least in part, explain the diverse broadband and line-emission/absorption properties seen in TDEs \citep{2018ApJ...859L..20D,2020MNRAS.494.4914P}.

\cite{2020MNRAS.494.4914P} simulated spectra with broad emission lines (BELs) and BALs with a disk and wind model for TDEs. In particular, they rendered spectra that consist of purely broad emission lines (BELs) at sightlines that do not intersect the disk wind. While viewing angle may be responsible for the dichotomy of BELs and BALs in TDE spectra, it does not  explain how one object can transition from BEL to BAL, as seen in AT~2018zr during the first 60 days \citep{2019ApJ...879..119H}.
The SLSN-like early-time spectrum of AT~2019qiz further calls for a time-dependent, multizone wind model to be considered in future TDE simulations.
Including AT~2019qiz, the high occurrence of BALs in TDE spectra (c.f. QSOs) continues to support a wide-angle wind geometry in TDEs \citep{2019ApJ...879..119H,2020MNRAS.494.4914P}.

\subsection{Origin of the Multicomponent H$\alpha$ Emission}
\begin{figure}
\centering
\includegraphics[width=3.5in, angle=0]{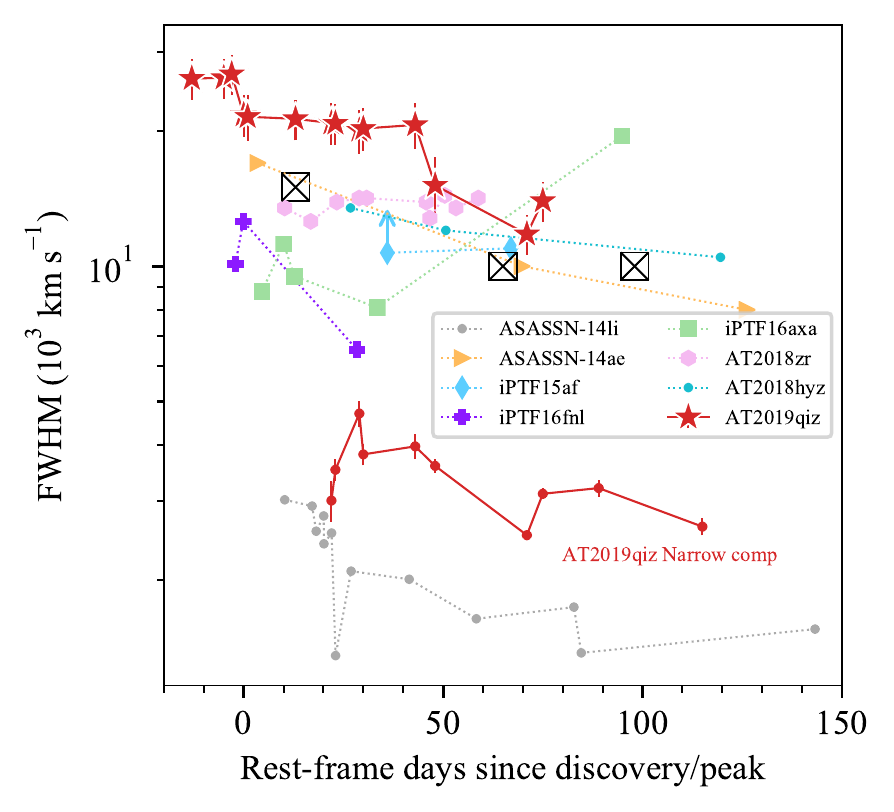}
\caption{The evolution of the AT~2019qiz \Ha\ FWHM compared with other TDEs. The FWHM of the broad \Ha\ component is shown with red stars while the FWHM of the narrow \Ha\ component is shown with red circles. Crossed boxes mark the BAL velocity at three {\it HST} epochs. It can be seen that the timing of the decrease in the broad \Ha\ width is consistent with the change in outflow speed probed by the BALs.
  \label{fig:FWHM_evo}}
\end{figure}

The accretion disk, outflow, and bound stellar debris are all possible production sites of \Ha\ emission in a TDE. At a glace, the broad \Ha\ component in AT~2019qiz may be reminiscent of a disk line profile with an intermediate inclination angle ($i$) such that the peaks corresponding to material moving away and toward the observer are not resolved \citep[e.g.,][]{2019ApJ...880..120H,2017MNRAS.472L..99L}. If due to rotation, the measured separation of $\sim15,000$\,\kms\ between the two fitted peaks in the pre-peak epochs implies a disk size of $400(\frac{\mathrm{sin}\,i}{0.5})^2$\,$R_g$. However, the fast evolution of this broad component, with FWHM shrinking by as much as $\sim50$\% in roughly 80 days, is hard to reconcile with the disk origin since changes in the disk geometry are expected to occur on longer timescales. In the double-peaked TDE AT~2018hyz, the \Ha\ line widths remained roughly constant for at least a duration of 100~days \citep[\autoref{fig:FWHM_evo};][]{2020apJ...903...31H}.

The width of the broad Balmer emission in TDEs is typically on the order of $10^4$~\kms. This value is comparable to the expected outflow velocity \citep[e.g.,][]{2009MNRAS.400.2070S,2016MNRAS.461..948M} in a TDE and therefore hints on the association. \cite{2018ApJ...855...54R} first proposed that asymmetric emission lines may result from an optically thick, outflowing gas that is expected to be common in TDEs. The resulting line lacks the blueshifted absorption portion of a P~Cygni line profile when the excitation temperature of the line is higher than the brightness temperature of the photosphere. Indeed, progressively more TDEs are found to have emission-line shapes deviating from a simple Gaussian (e.g., ASASSN-14ae, AT2018~zr, AT2018~hyz). \cite{2019ApJ...879..119H} found a good fit to the flat-topped \Ha\ emission in AT2018~zr with this model, though an elliptical disk within a certain parameter space also produces such line shapes \citep{2019ApJ...880..120H}. \citetalias{2020mnRas.499..482N} explored this possibility by comparing the \cite{2018ApJ...855...54R} model with the spectra of AT~2019qiz. In summary, \citetalias{2020mnRas.499..482N} found the model with an outflow velocity of 5000\,\kms\ to be the best match to the \Ha\ line profile at earlier and later epochs in AT~2019qiz. However, at the epochs where the \Ha\ emission is the strongest, the \cite{2018ApJ...855...54R} model fails to replicate the strong red wing of the line profile. The discrepancy may be resolved by adopting a higher opacity, changing the photosphere radius, or allowing for asymmetries in the outflow.

The potential correlation between the {\it HST} UV spectral features and the broad \Ha\ component in AT~2019qiz could lend further support to the outflow origin of the the broad \Ha\ emission. We observed a reduction in the velocity of both the HiBALs in the UV and in the broad \Ha\ component in the optical (\autoref{fig:FWHM_evo}). The optical spectra with a higher cadence than the {\it HST} UV spectra find the FWHM to decrease from $\sim22,000$~\kms\ on $\Delta t=13$ days to $\lesssim15,000$~\kms\ on $\Delta t\geq48$~days. This change is well-reflected in the $\Delta t=65$~days {\it HST} spectra, which decelerated by $\sim5000$~\kms\ since the first {\it HST} observation on $\Delta t=13$~days.

\subsection{Origin of the Narrow UV Absorption Lines}

As mentioned, we are unable to constrain the variability in the narrow absorption lines confidently owing to the lower S/N at the later {\it HST} epochs. However, the FUV spectrum from the second {\it HST} epoch does have sufficient S/N to reveal stronger absorption lines with $W_r\gtrsim 1.0$. We find these lines to have a comparable strength in the day 13 and day 65 spectra.

\begin{figure*}
\centering
\includegraphics[width=7in, angle=0]{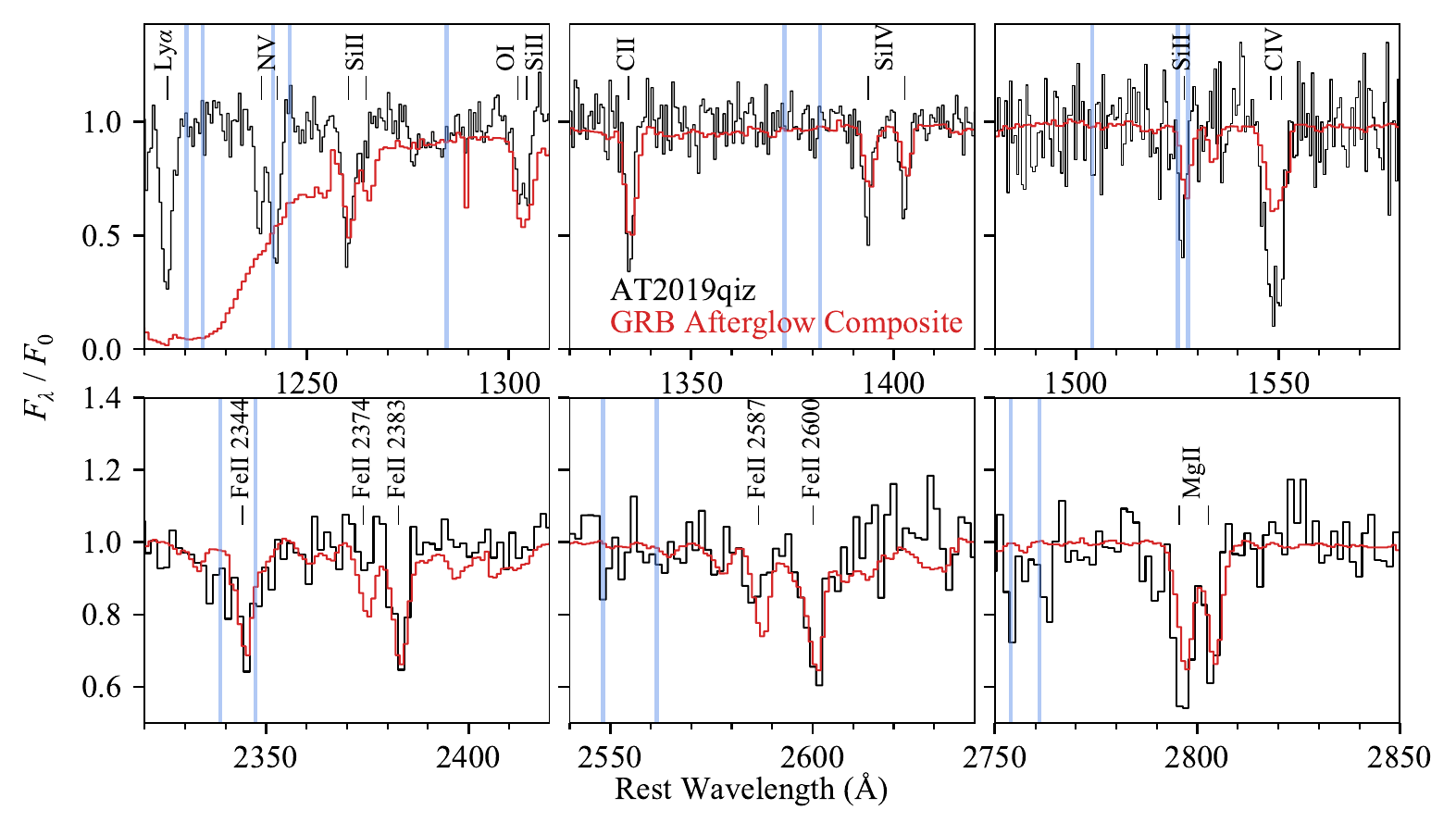}
\caption{Comparison of the normalized spectrum of AT2019~qiz (black) with the GRB composite from \cite{2011ApJ...727...73C} (red). We indicate the wavelength for different transitions in the host-galaxy rest frame with black labels and possible contamination by foreground absorption ($z\approx0$) with blue vertical lines.
  \label{fig:grb}}
\end{figure*}

GRB afterglow spectra have long been exploited to study the intervening interstellar medium (ISM) and circumgalactic medium (CGM) along the line of sight \citep[e.g.,][]{2007ApJ...666..267P,2019ApJ...884...66G}.
To shine light on the origin of the narrow UV absorption lines in AT~2019qiz, we compare its spectrum with that of the GRB afterglow composite \citep[\autoref{fig:grb};][]{2011ApJ...727...73C}. In AT~2019qiz, we derived $N$(\ion{H}{1}) $=5\times10^{17}$\,cm$^{-2}$ from the CoG analysis. This column density is consistent with a Lyman-limit system, where the cloud starts to be dense enough to shield itself against the EUV background and remain neutral. GRB afterglow spectra, on the other hand, are often associated with a damped Ly$\alpha$ (DLA) system with a high neutral hydrogen column $\textrm{log}(N(\textrm{\ion{H}{1}})) > 20.3$~cm$^{-2}$. The DLA line is dominated by the Voigt profile that is very different from the Gaussian-like Ly$\alpha$ line profile in AT~2019qiz (\autoref{fig:grb}).

We observe stronger \SiIVs\ and \CIVs\ absorption in AT~2019qiz than in the GRB composite. The strengths of the low-ionization species (\ion{Fe}{2} and \ion{Mg}{2}) are comparable to those in GRB afterglows. While the \NV\ absorption lines are well-detected in AT~2019qiz, they are harder to identify in GRB spectra owing to blending with the red wing of Ly$\alpha$. The high ionization potential of N$^{4+}$ (77~eV) makes it difficult to produce in stellar radiation fields. Galactic halos and disks do have \NVs\ absorption, but none of them show a column density $>10^{14.5}$~cm$^{-2}$ (Prochaska et al. 2008, and references therein).  Our measured N$^{4+}$ column density of $\gtrsim10^{15}$~cm$^{-2}$ thus likely indicates a hard ionizing spectrum, which can be attributed to the TDE accretion disk or in part the underlying weak AGN. The strong ionizing source may also explain the stronger \SiIVs\ and \CIVs\ $W_r$ in AT~2019qiz than in the GRB composite. These high-ionization lines are likely to have formed in regions close to the SMBH rather than in the ISM. Our {\it HST} spectra cannot resolve fine-structure transitions such as \ion{Si}{2}$^*$ and \ion{Fe}{2}$^*$ in AT~2019qiz. In GRB afterglows, these fine-structure lines are thought to be produced by UV pumping and uniquely trace gas in the vicinity (100--1000\,pc) of the burst \citep{2006ApJ...650..272P}. We expect these lines to be detected in high S/N, high-resolution TDE spectra since TDEs can also create a temporary UV radiation field.

Given the redshift of the absorption lines being close to the host redshift, we consider both circumnuclear gas and the bound stellar debris as the candidate absorbing system. In ASASSN-14li, \cite{Cenko2016} also detected narrow absorption lines with a width of $\sim500$~\kms\ comparable to that in AT~2019qiz. However, the absorption lines in ASASSN-14li are blueshifted by 250--400\,\kms\ while the lines in AT~2019qiz do not appear to have a systemic velocity shift with respect to the host. The abundance pattern in ASASSN-14li is also very different. The low neutral hydrogen column of $\textrm{log}(N(\textrm{\ion{H}{1}})) < 14.2$~cm$^{-2}$ is somewhat expected in host galaxies with an old stellar population \citep{Cenko2016}. However, ASASSN-14li lacks the low-ionization \MgII\ absorption that is commonly observed in cold ISM in the galaxy while having well-detected high-ionization lines. Together with the blueshift velocity being consistent with the outflow velocity measured from X-ray spectroscopy, \cite{Cenko2016} suggest that the low-velocity absorber is more likely the bound stellar debris on an elliptical orbit \citep{2015Natur.526..542M,Cenko2016}.

The absorber in AT~2019qiz has several properties different from that in ASASSN-14li. The low-ionization species, such as \ion{Fe}{2} and \ion{Mg}{2}, are readily detected in AT~2019qiz with $W_r$ comparable to those observed in GRB afterglow and QSO spectra, where the absorbing systems have been associated with galaxies. The likely lack of variability also counters the scenario in which the absorbing gas is the bound stellar debris. Clumpiness or bulk motion of the debris stream could easily alter the gas density and therefore lead to variability in the absorption lines on the fallback timescale $t_\mathrm{fallback}\approx40$ days. Therefore, we conclude that the absorption lines in AT2019~qiz are more likely to be probing the host ISM and circumnuclear gas instead of the stellar debris as in ASASSN-14li.

\section{Conclusions} \label{sec:conclusions}
The early classification of AT~2019qiz and its proximity allow us to perform a detailed, multiwavelength study of a TDE. Our UV and optical follow-up observations in both photometric and spectroscopic modes offer new insights for optically-selected TDEs. AT~2019qiz is located in a galaxy with a weak AGN, as evidenced by the line ratios in the BPT diagram. The evolution timescale and luminosity of AT~2019qiz are fast and faint, making it an intermediate event between iPTF16fnl and the rest of the TDE population.

Both the photosphere evolution and the BALs in the \textit{HST} spectra strongly corroborate the outflow scenario in TDEs. We find the photosphere to be expanding at a constant velocity $v \approx 2,700$~\kms\ before reaching $t_{\rm peak}$. The early luminosity evolution in AT~2019qiz follows a power law ($\propto t^2$), consistent with the predication of an expanding fireball. The BALs are detected at all three \textit{HST} epochs with a decrease in outflow velocity from 15,000\,\kms\ to 10,000\,\kms. This deceleration is expected as the mass fallback rate declines at later times.

The UV spectrum at the first \textit{HST} epoch is unlike that of all the TDEs observed in the past, as it contains both HiBAL and FeLoBAL features such as \ion{Al}{3} and \ion{Fe}{2}. The UV spectrum of AT~2019qiz resembles that of the SLSN Gaia16apd at maximum light more than that of any known FeLoBAL QSO. However, the broad \MgII\ doublet, which is readily found in both Gaia16apd and FeLoBAL QSOs, is not detected in AT~2019qiz. The disappearance of the FeLoBAL features at the later two \textit{HST} epochs of AT~2019qiz is likely due to the thinning of the X-ray/EUV shielding material as the mass fallback/accretion rate decreases with time.

The detection of \HeII\ and the Balmer emission since the beginning of the monitoring up to day 89 makes AT~2019qiz a TDE-Bowen in the spectral classification scheme proposed by \cite{2020arXiv200805461V}. We carefully studied the \Ha\ emission and found the line profile to be characterized by a very broad component with FWHM $\gtrsim10^4$~\kms\ and a narrower component with FWHM $\approx 3000$~\kms\ that only appeared after $\Delta t\gtrsim22$ days. The broad \Ha\ component narrows with time as observed in other TDEs and becomes progressively more symmetric around the rest wavelength of \Ha. We inferred that the broad \Ha\ evolution is most likely to be driven by the outflow given the similar value of the \Ha\ line width and the BAL velocity, and the fact that the timing of the decrease in broad \Ha\ width is seen to be coincident with the decrease in wind velocity probed by the BALs.

We also analyzed the narrow absorption lines in the first \textit{HST} spectrum and derived ionic column densities for several species using the CoG method. The measured Ly$\alpha$ $W_r$ corresponds to a Lyman-limit system, where the high-ionization species have stronger $W_r$ and the low-ionization lines have similar $W_r$ compared with those in GRB afterglows. We concluded that the narrow absorption lines are more likely to be probing the gas in the host galaxy rather than the bound stellar debris as in the case of ASASSN-14li. The UV spectra of AT~2019qiz demonstrate the potential of using TDEs to probe circumnuclear gas and the ISM in TDE host galaxies. Future high-S/N and high-resolution UV spectra will allow us to resolve the fine-structure lines produced by UV pumping in TDEs and study the gas in the vicinity of the SMBHs.

\acknowledgments
We thank anonymous referee for helpful suggestions that greatly improved the quality of the paper. T.H.\ is grateful to Ari Laor, Alexei Baskin, and Frederick Hamann for generously sharing their insightful thoughts on the UV spectral signatures in AT~2019qiz and providing BALQSO templates for comparison. She also thanks Dan Perley for helpful discussion on the similarities between AT~2019qiz and Gaia16apd. T.H.\ thanks Elizabeth Nance and John Debes for their help with scheduling the {\it HST} ToO observations.

The UCSC transient team is supported in part by National Science Foundation (NSF) grant AST-1518052, National Aeronautics and Space Administration (NASA) {\it Swift} grant 80NSSC19K1386, the Gordon \& Betty Moore Foundation, the Heising-Simons Foundation, and by a fellowship to R.J.F. from the David and Lucile Packard Foundation.  K.A.A., E.R.-R., and B.M.\ are supported by the Danish National Research Foundation (DNRF132), the Heising-Simons Foundation, and NSF grant AST-161588. J.L.D.\ is  supported  by  the  GRF  grant  from the Hong Kong government under HKU 27305119. M.R.S.\ is supported by the NSF Graduate Research Fellowship Program Under grant 1842400. A.V.F.'s group is grateful for generous financial assistance from the Christopher R. Redlich Fund, the TABASGO Foundation, and the Miller Institute for Basic Research in Science (U.C. Berkeley; A.V.F. is a Miller Senior Fellow). Support for T.W.-S.H.\ was provided by NASA through NASA Hubble Fellowship grant HST-HF2-51458.001-A awarded by the Space Telescope Science Institute (STScI), which is operated by the Association of Universities for Research in Astronomy (AURA), Inc., under NASA contract NAS 5-26555. Parts of this research were supported by the Australian Research Council Centre of Excellence for All Sky Astrophysics in 3 Dimensions (ASTRO 3D), through project number CE170100013. Research at Lick Observatory is partially supported by a generous gift from Google.

This work is based on observations made with the NASA/ESA {\it Hubble Space Telescope} under program number GO-16026. Support for program GO-16026 was provided by NASA through a grant from STScI, which is operated by AURA, Inc., under NASA contract NAS 5-26555. This work makes use of observations from the Las Cumbres Observatory global telescope network following the approved program 2019B-0363.
Some of the data presented herein were obtained at the W.\ M.\ Keck Observatory, which is operated as a scientific partnership among the California Institute of Technology, the University of California, and NASA. The Observatory was made possible by the generous financial support of the W.\ M.\ Keck Foundation.  The authors wish to recognize and acknowledge the very significant cultural role and reverence that the summit of Maunakea has always had within the indigenous Hawaiian community.  We are most fortunate to have the opportunity to conduct observations from this mountain.

Based in part on observations obtained at the Southern Astrophysical Research (SOAR) telescope, which is a joint project of the Minist\'{e}rio da Ci\^{e}ncia, Tecnologia, Inova\c{c}\~{o}es e Comunica\c{c}\~{o}es (MCTIC) do Brasil, the U.S. National Optical Astronomy Observatory (NOAO), the University of North Carolina at Chapel Hill (UNC), and Michigan State University (MSU).
This work includes data obtained with the Swope Telescope
at Las Campanas Observatory, Chile, as part of the Swope Time Domain Key Project (PI Piro, Co-Is Burns, Cowperthwaite, Dimitriadis, Drout, Foley, French, Holoien, Hsiao, Kilpatrick, Madore, Phillips, and Rojas-Bravo). The authors thank Swope Telescope observers Jorge Anais Vilchez, Abdo Campillay, Yilin Kong Riveros, and Natalie Ulloa for collecting data presented in this paper.

\software{photpipe imaging and photometry pipeline \citep{Rest2005,Rest2014}, hotpants \citep{HOTPANTS}, DoPhot \citep{Schechter93}, PyRAF (Science Software Branch at STScI 2012), pPXF \citep{2017MNRAS.466..798C}, Prospector \citep{2017ApJ...837..170L}, HEASoft \citep{2014ascl.soft08004N}, MOSFiT \citep{2018ApJS..236....6G,Mockler2019}}

\bibliography{tde}{}
\bibliographystyle{aasjournal}



\appendix

\restartappendixnumbering
\renewcommand{\thetable}{A\arabic{table}}
\startlongtable
\begin{deluxetable}{lccc}
\tablewidth{0pt}
\tablecolumns{4}
\tablecaption{Photometric data of AT~2019qiz \label{tab: phot}}
\tabletypesize{\footnotesize}
\tablehead{\colhead{MJD} & \colhead{Magnitude} & \colhead{Filter} & \colhead{Telescope}}
\startdata
\input{phot.tex}
\enddata
\tablecomments{All the data presented in this table have been corrected for Galactic extinction. We did not perform host subtraction on the \swift\ UV data since the host contribution is negligible. On the other hand, the ZTF, LCO, and Swope data are all host-subtracted.}
\end{deluxetable}

\end{document}

%% file: line_names.tex
\newcommand{\La}{\ifmmode {\rm Ly}\alpha \else Ly$\alpha$\fi}
\newcommand{\Ka}{\ifmmode {\rm K}\alpha \else K$\alpha$\fi}
\newcommand{\Lb}{\ifmmode {\rm L}\beta \else L$\beta$\fi}
\newcommand{\Ha}{\ifmmode {\rm H}\alpha \else H$\alpha$\fi}
\newcommand{\Hb}{\ifmmode {\rm H}\beta \else H$\beta$\fi}
\newcommand{\Hgamma}{\ifmmode {\rm H}\gamma \else H$\gamma$\fi}
\newcommand{\Hdelta}{\ifmmode {\rm H}\delta \else H$\delta$\fi}
\newcommand{\Pa}{\ifmmode {\rm P}\alpha \else P$\alpha$\fi}
\newcommand{\CaII}{\ifmmode {\rm Ca}\,{\sc ii]}\,\lambda\lambda3934, 3969
    \else Ca\,{\sc ii]}\,$\lambda\lambda3934, 3969$\fi}
\newcommand{\CIIIb}{\ifmmode {\rm C}\,{\sc iii]}\,\lambda1909
    \else C\,{\sc iii]}\,$\lambda1909$\fi}
\newcommand{\CIV}{\ifmmode {\rm C}\,{\sc iv}\,\lambda1548
    \else C\,{\sc iv}\,$\lambda\lambda1548, 1551$\fi}
\newcommand{\CIVs}{\ifmmode {\rm C}\,{\sc iv}
    \else C\,{\sc iv}\fi}

\newcommand{\OVI}{\ifmmode {\rm O}\,{\sc vi}\,\lambda1035
    \else O\,{\sc vi}\,$\lambda1035$\fi}
\newcommand{\HI}{\ifmmode {\rm H}\,{\sc i}
    \else H\,{\sc i}\fi}
\newcommand{\HeIs}{\ifmmode {\rm He}\,{\sc i}\,^*\lambda3889
    \else He\,{\sc i}\,$^*\lambda3889$\fi}
\newcommand{\HeIss}{\ifmmode {\rm He}\,{\sc i}\,^*\lambda10830
    \else He\,{\sc i}\,$^*\lambda10830$\fi}
\newcommand{\HeI}{\ifmmode {\rm He}\,{\sc i}\,^*\lambda5876
    \else He\,{\sc i}\,$^*\lambda5876$\fi}
\newcommand{\HeII}{\ifmmode {\rm He}\,{\sc ii}\,\lambda4686
    \else He\,{\sc ii}\,$\lambda4686$\fi}
\newcommand{\HeIIa}{\ifmmode {\rm He}\,{\sc ii}\,\lambda1640
    \else He\,{\sc ii}\,$\lambda1640$\fi}
\newcommand{\HeIIb}{\ifmmode {\rm He}\,{\sc ii}\,\lambda1085
    \else He\,{\sc ii}\,$\lambda1085$\fi}
\newcommand{\CII}{\ifmmode {\rm C}\,{\sc ii}\,\lambda1335
    \else C\,{\sc ii}\,$\lambda1335$\fi}
\newcommand{\CIII}{\ifmmode {\rm C}\,{\sc iii}\,\lambda977
    \else C\,{\sc iii}\,$\lambda977$\fi}

\newcommand{\bOIIIb}{\ifmmode {\rm [O}\,{\sc iii]}\,\lambda5007
    \else [O\,{\sc iii]}\,$\lambda5007$\fi}
\newcommand{\OIIIb}{\ifmmode {\rm O}\,{\sc iii]}\,\lambda1663
    \else O\,{\sc iii]}\,$\lambda1663$\fi}
\newcommand{\OVIII}{\ifmmode {\rm O}\,{\sc viii}\,653~{\rm eV}
    \else O\,{\sc viii}\,$653~{\rm eV}$\fi}
\newcommand{\OVII}{\ifmmode {\rm O}\,{\sc vii}\,568~eV
    \else O\,{\sc vii}\,$568~{\rm eV}$\fi}

\newcommand{\OIVb}{\ifmmode {\rm O}\,{\sc iv]}\,\lambda1402
    \else O\,{\sc iv]}\,$\lambda1402$\fi}
\newcommand{\bOIIb}{\ifmmode {\rm [O}\,{\sc ii]}\,\lambda3727
    \else [O\,{\sc ii]}\,$\lambda3727$\fi}
\newcommand{\bOIb}{\ifmmode {\rm [O}\,{\sc i]}\,\lambda6300
    \else [O\,{\sc i]}\,$\lambda6300$\fi}
\newcommand{\OI}{\ifmmode {\rm [O}\,{\sc i]}\,\lambda1304
    \else [O\,{\sc i]}\,$\lambda1304$\fi}
\newcommand{\NII}{\ifmmode {\rm N}\,{\sc ii}\,\lambda1084
    \else N\,{\sc ii}\,$\lambda1084$\fi}
\newcommand{\NIII}{\ifmmode {\rm N}\,{\sc iii}\,\lambda990
    \else N\,{\sc iii}\,$\lambda990$\fi}
\newcommand{\NIIIb}{\ifmmode {\rm N}\,{\sc iii]}\,\lambda1750
    \else N\,{\sc iii]}\,$\lambda1750$\fi}
\newcommand{\NIVb}{\ifmmode {\rm N}\,{\sc iv]}\,\lambda1486
    \else N\,{\sc iv]}\,$\lambda1486$\fi}
\newcommand{\NV}{\ifmmode {\rm N}\,{\sc v}\,\lambda\lambda1238, 1242
    \else N\,{\sc v}\,$\lambda\lambda1238, 1242$\fi}
\newcommand{\NVs}{\ifmmode {\rm N}\,{\sc v}
    \else N\,{\sc v}\fi}
\newcommand{\MgII}{\ifmmode {\rm Mg}\,{\sc ii}\,\lambda\lambda2796, 2803
      \else Mg\,{\sc ii}\,$\lambda\lambda2796, 2803$\fi}
\newcommand{\CVI}{\ifmmode {\rm C}\,{\sc vi}\,368~eV
      \else C\,{\sc vi}\,368~eV\fi}
\newcommand{\SiIV}{\ifmmode {\rm Si}\,{\sc iv}\,\lambda1397
    \else Si\,{\sc iv}\,$\lambda1397$\fi}
\newcommand{\SiIVs}{\ifmmode {\rm Si}\,{\sc iv}
    \else Si\,{\sc iv}\fi}
\newcommand{\bFeXb}{\ifmmode {\rm [Fe}\,{\sc x]}\,\lambda6734
      \else [Fe\,{\sc x]}\,$\lambda6734$\fi}
\newcommand{\MgX}{\ifmmode {\rm Mg}\,{\sc x}\,\lambda615
      \else Mg\,{\sc x}\,$\lambda615$\fi}
\newcommand{\MgXI}{\ifmmode {\rm Mg}\,{\sc xi}\,1.34~keV
      \else Mg\,{\sc xi}\,1.34~keV\fi}
\newcommand{\MgXII}{\ifmmode {\rm Mg}\,{\sc xii}\,1.47~keV
    \else Mg\,{\sc xii}\,1.47~keV\fi}
\newcommand{\bNeVb}{\ifmmode {\rm [Ne}\,{\sc v]}\,\lambda3426
    \else [Ne\,{\sc v]}\,$\lambda3426$\fi}
\newcommand{\NeVIII}{\ifmmode {\rm Ne}\,{\sc viii}\,\lambda774
    \else Ne\,{\sc viii}\,$\lambda774$\fi}
\newcommand{\SiVIIa}{\ifmmode {\rm Si}\,{\sc vii}\,\lambda70
    \else Si\,{\sc vii}\,$\lambda70$\fi}
\newcommand{\NeVIIa}{\ifmmode {\rm Ne}\,{\sc vii}\,\lambda88
    \else Ne\,{\sc vii}\,$\lambda88$\fi}
\newcommand{\NeVIIIa}{\ifmmode {\rm Ne}\,{\sc viii}\,\lambda88
    \else Ne\,{\sc viii}\,$\lambda88$\fi}
\newcommand{\NeIX}{\ifmmode {\rm Ne}\,{\sc ix}\,915~eV
    \else Ne\,{\sc ix}\,915~eV\fi}
\newcommand{\NeX}{\ifmmode {\rm Ne}\,{\sc x}\,1.02~keV
    \else Ne\,{\sc x}\,1.02~keV\fi}
\newcommand{\SiII}{\ifmmode {\rm Si}\,{\sc ii}\,\lambda1265
    \else Si\,{\sc ii}\,$\lambda1265$\fi}
\newcommand{\SiIII}{\ifmmode {\rm Si}\,{\sc iii}\,\lambda1206
    \else Si\,{\sc iii}\,$\lambda1206$\fi}
\newcommand{\SiXII}{\ifmmode {\rm Si}\,{\sc xii}\,\lambda506
      \else Si\,{\sc xii}\,$\lambda506$\fi}
\newcommand{\SiXIII}{\ifmmode {\rm Si}\,{\sc xiii}\,1.85~keV
    \else Si\,{\sc xiii}\,1.85~keV\fi}
\newcommand{\SiXIV}{\ifmmode {\rm Si}\,{\sc xiv}\,2.0~keV
    \else Si\,{\sc xiv}\,2.0~keV\fi}
\newcommand{\SXV}{\ifmmode {\rm S}\,{\sc xv}\,2.45~keV
    \else S\,{\sc xv}\,2.45~keV\fi}
\newcommand{\SXVI}{\ifmmode {\rm S}\,{\sc xvi}\,2.62~keV
    \else S\,{\sc xvi}\,2.62~keV\fi}
\newcommand{\ArXVII}{\ifmmode {\rm Ar}\,{\sc xvii}\,3.10~keV
    \else Ar\,{\sc xvii}\,3.10~keV\fi}
\newcommand{\ArXVIII}{\ifmmode {\rm Ar}\,{\sc xviii}\,3.30~keV
    \else Ar\,{\sc xviii}\,3.30~keV\fi}
\newcommand{\FeIII}{\ifmmode {\rm Fe}\,{\sc iii}
    \else Fe\,{\sc iii}\fi}
\newcommand{\FeIXVI}{\ifmmode {\rm Fe}\,{\sc 1-16}\,6.4~keV
    \else Fe\,{\sc 1-16}\,6.4~keV\fi}
\newcommand{\FeXVIIXXIII}{\ifmmode {\rm Fe}\,{\sc 17-23}\,6.5~keV
    \else Fe\,{\sc 17-23}\,6.5~keV\fi}
\newcommand{\FeXXV}{\ifmmode {\rm Fe}\,{\sc xxv}\,6.7~keV
    \else Fe\,{\sc xxv}\,6.7~keV\fi}
\newcommand{\FeXXVI}{\ifmmode {\rm Fe}\,{\sc xxvi}\,6.96~keV
    \else Fe\,{\sc xxvi}\,6.96~keV\fi}
\newcommand{\FeLa}{\ifmmode {\rm Fe}\,{\sc L}\,0.7-0.8~keV
    \else Fe\,{\sc L}\,0.7-0.8~keV\fi}
\newcommand{\FeLb}{\ifmmode {\rm Fe}\,{\sc L}\,1.03-1.15~keV
    \else Fe\,{\sc L}\,1.03-1.15~keV\fi}
\newcommand{\AlIII}{\ifmmode {\rm Al}\,{\sc iii}\,\lambda1857
    \else Al\,{\sc iii}\,$\lambda1857$\fi}

%% file: phot.tex
58753.059 & 16.10 $\pm$ 0.04 & UVW2 & \textsl{Swift} \\
58755.388 & 16.01 $\pm$ 0.05 & UVW2 & \textsl{Swift} \\
58756.436 & 15.79 $\pm$ 0.05 & UVW2 & \textsl{Swift} \\
58763.212 & 15.16 $\pm$ 0.05 & UVW2 & \textsl{Swift} \\
58766.074 & 15.34 $\pm$ 0.04 & UVW2 & \textsl{Swift} \\
58769.324 & 15.60 $\pm$ 0.04 & UVW2 & \textsl{Swift} \\
58772.787 & 15.82 $\pm$ 0.11 & UVW2 & \textsl{Swift} \\
58778.548 & 16.27 $\pm$ 0.05 & UVW2 & \textsl{Swift} \\
58781.681 & 16.85 $\pm$ 0.06 & UVW2 & \textsl{Swift} \\
58784.603 & 16.61 $\pm$ 0.05 & UVW2 & \textsl{Swift} \\
58787.063 & 16.83 $\pm$ 0.06 & UVW2 & \textsl{Swift} \\
58790.640 & 16.99 $\pm$ 0.06 & UVW2 & \textsl{Swift} \\
58796.034 & 17.10 $\pm$ 0.13 & UVW2 & \textsl{Swift} \\
58799.813 & 17.14 $\pm$ 0.07 & UVW2 & \textsl{Swift} \\
58802.925 & 17.24 $\pm$ 0.07 & UVW2 & \textsl{Swift} \\
58805.257 & 17.41 $\pm$ 0.08 & UVW2 & \textsl{Swift} \\
58809.101 & 17.50 $\pm$ 0.07 & UVW2 & \textsl{Swift} \\
58815.346 & 17.58 $\pm$ 0.07 & UVW2 & \textsl{Swift} \\
58820.980 & 17.99 $\pm$ 0.09 & UVW2 & \textsl{Swift} \\
58825.841 & 18.01 $\pm$ 0.09 & UVW2 & \textsl{Swift} \\
58836.845 & 18.47 $\pm$ 0.14 & UVW2 & \textsl{Swift} \\
58847.951 & 18.59 $\pm$ 0.11 & UVW2 & \textsl{Swift} \\
58852.583 & 18.60 $\pm$ 0.11 & UVW2 & \textsl{Swift} \\
58857.113 & 18.88 $\pm$ 0.13 & UVW2 & \textsl{Swift} \\
58862.350 & 18.92 $\pm$ 0.13 & UVW2 & \textsl{Swift} \\
58867.939 & 19.09 $\pm$ 0.19 & UVW2 & \textsl{Swift} \\
58929.624 & 19.38 $\pm$ 0.18 & UVW2 & \textsl{Swift} \\
58753.064 & 16.17 $\pm$ 0.05 & UVM2 & \textsl{Swift} \\
58755.392 & 16.05 $\pm$ 0.05 & UVM2 & \textsl{Swift} \\
58756.439 & 15.79 $\pm$ 0.05 & UVM2 & \textsl{Swift} \\
58763.214 & 15.08 $\pm$ 0.05 & UVM2 & \textsl{Swift} \\
58766.078 & 15.08 $\pm$ 0.04 & UVM2 & \textsl{Swift} \\
58769.328 & 15.26 $\pm$ 0.04 & UVM2 & \textsl{Swift} \\
58778.551 & 15.89 $\pm$ 0.05 & UVM2 & \textsl{Swift} \\
58781.684 & 16.31 $\pm$ 0.05 & UVM2 & \textsl{Swift} \\
58784.606 & 16.24 $\pm$ 0.05 & UVM2 & \textsl{Swift} \\
58787.066 & 16.39 $\pm$ 0.05 & UVM2 & \textsl{Swift} \\
58790.643 & 16.57 $\pm$ 0.05 & UVM2 & \textsl{Swift} \\
58799.816 & 16.82 $\pm$ 0.06 & UVM2 & \textsl{Swift} \\
58802.928 & 16.88 $\pm$ 0.06 & UVM2 & \textsl{Swift} \\
58805.259 & 17.02 $\pm$ 0.07 & UVM2 & \textsl{Swift} \\
58809.104 & 17.26 $\pm$ 0.07 & UVM2 & \textsl{Swift} \\
58815.349 & 17.52 $\pm$ 0.07 & UVM2 & \textsl{Swift} \\
58820.983 & 17.62 $\pm$ 0.08 & UVM2 & \textsl{Swift} \\
58825.844 & 17.96 $\pm$ 0.09 & UVM2 & \textsl{Swift} \\
58836.847 & 18.07 $\pm$ 0.11 & UVM2 & \textsl{Swift} \\
58847.955 & 18.39 $\pm$ 0.10 & UVM2 & \textsl{Swift} \\
58852.587 & 18.57 $\pm$ 0.10 & UVM2 & \textsl{Swift} \\
58857.117 & 18.61 $\pm$ 0.11 & UVM2 & \textsl{Swift} \\
58862.353 & 18.69 $\pm$ 0.11 & UVM2 & \textsl{Swift} \\
58867.941 & 18.98 $\pm$ 0.18 & UVM2 & \textsl{Swift} \\
58910.346 & 18.95 $\pm$ 0.09 & UVM2 & \textsl{Swift} \\
58934.139 & 19.68 $\pm$ 0.30 & UVM2 & \textsl{Swift} \\
58753.054 & 16.31 $\pm$ 0.05 & UVW1 & \textsl{Swift} \\
58755.385 & 16.16 $\pm$ 0.05 & UVW1 & \textsl{Swift} \\
58756.433 & 15.97 $\pm$ 0.05 & UVW1 & \textsl{Swift} \\
58759.755 & 15.53 $\pm$ 0.05 & UVW1 & \textsl{Swift} \\
58759.768 & 15.47 $\pm$ 0.05 & UVW1 & \textsl{Swift} \\
58760.637 & 15.44 $\pm$ 0.05 & UVW1 & \textsl{Swift} \\
58763.211 & 15.30 $\pm$ 0.05 & UVW1 & \textsl{Swift} \\
58766.071 & 15.31 $\pm$ 0.05 & UVW1 & \textsl{Swift} \\
58769.321 & 15.43 $\pm$ 0.05 & UVW1 & \textsl{Swift} \\
58771.188 & 15.56 $\pm$ 0.04 & UVW1 & \textsl{Swift} \\
58771.201 & 15.55 $\pm$ 0.04 & UVW1 & \textsl{Swift} \\
58772.784 & 15.63 $\pm$ 0.05 & UVW1 & \textsl{Swift} \\
58778.545 & 15.97 $\pm$ 0.05 & UVW1 & \textsl{Swift} \\
58781.678 & 16.24 $\pm$ 0.06 & UVW1 & \textsl{Swift} \\
58784.600 & 16.28 $\pm$ 0.06 & UVW1 & \textsl{Swift} \\
58787.060 & 16.46 $\pm$ 0.06 & UVW1 & \textsl{Swift} \\
58790.637 & 16.57 $\pm$ 0.06 & UVW1 & \textsl{Swift} \\
58796.030 & 16.72 $\pm$ 0.06 & UVW1 & \textsl{Swift} \\
58799.810 & 16.83 $\pm$ 0.07 & UVW1 & \textsl{Swift} \\
58802.922 & 16.99 $\pm$ 0.07 & UVW1 & \textsl{Swift} \\
58805.255 & 17.21 $\pm$ 0.09 & UVW1 & \textsl{Swift} \\
58809.098 & 17.24 $\pm$ 0.08 & UVW1 & \textsl{Swift} \\
58815.343 & 17.40 $\pm$ 0.08 & UVW1 & \textsl{Swift} \\
58820.977 & 17.63 $\pm$ 0.09 & UVW1 & \textsl{Swift} \\
58825.838 & 17.60 $\pm$ 0.09 & UVW1 & \textsl{Swift} \\
58836.844 & 18.02 $\pm$ 0.13 & UVW1 & \textsl{Swift} \\
58847.949 & 18.13 $\pm$ 0.11 & UVW1 & \textsl{Swift} \\
58852.580 & 18.21 $\pm$ 0.11 & UVW1 & \textsl{Swift} \\
58857.110 & 18.46 $\pm$ 0.13 & UVW1 & \textsl{Swift} \\
58862.347 & 18.44 $\pm$ 0.12 & UVW1 & \textsl{Swift} \\
58867.278 & 18.68 $\pm$ 0.14 & UVW1 & \textsl{Swift} \\
58899.463 & 18.70 $\pm$ 0.10 & UVW1 & \textsl{Swift} \\
58919.532 & 18.89 $\pm$ 0.14 & UVW1 & \textsl{Swift} \\
58751.989 & 15.98 $\pm$ 0.17 & u & Siding Spring 1m \\
58752.350 & 16.38 $\pm$ 0.17 & u & Swope \\
58753.268 & 16.38 $\pm$ 0.17 & u & Swope \\
58754.335 & 16.23 $\pm$ 0.17 & u & Swope \\
58754.759 & 15.98 $\pm$ 0.17 & u & Siding Spring 1m \\
58755.381 & 15.91 $\pm$ 0.17 & u & Swope \\
58757.979 & 15.48 $\pm$ 0.17 & u & Siding Spring 1m \\
58759.328 & 15.56 $\pm$ 0.17 & u & Swope \\
58760.366 & 15.43 $\pm$ 0.17 & u & Swope \\
58760.965 & 15.35 $\pm$ 0.17 & u & Siding Spring 1m \\
58761.336 & 15.34 $\pm$ 0.17 & u & Swope \\
58762.262 & 15.36 $\pm$ 0.17 & u & Swope \\
58763.961 & 15.24 $\pm$ 0.17 & u & Siding Spring 1m \\
58766.592 & 15.20 $\pm$ 0.17 & u & Siding Spring 1m \\
58769.655 & 15.32 $\pm$ 0.17 & u & Siding Spring 1m \\
58772.596 & 15.40 $\pm$ 0.17 & u & Siding Spring 1m \\
58774.316 & 15.59 $\pm$ 0.17 & u & Swope \\
58775.275 & 15.61 $\pm$ 0.17 & u & Swope \\
58776.322 & 15.82 $\pm$ 0.17 & u & Swope \\
58777.324 & 15.76 $\pm$ 0.17 & u & Swope \\
58778.242 & 15.82 $\pm$ 0.17 & u & Swope \\
58778.649 & 15.63 $\pm$ 0.17 & u & Siding Spring 1m \\
58781.729 & 15.71 $\pm$ 0.17 & u & Siding Spring 1m \\
58783.318 & 16.09 $\pm$ 0.17 & u & Swope \\
58784.269 & 16.09 $\pm$ 0.17 & u & Swope \\
58784.740 & 15.96 $\pm$ 0.17 & u & Siding Spring 1m \\
58785.287 & 16.21 $\pm$ 0.17 & u & Swope \\
58786.300 & 16.17 $\pm$ 0.17 & u & Swope \\
58787.249 & 16.31 $\pm$ 0.17 & u & Swope \\
58787.580 & 15.90 $\pm$ 0.18 & u & Siding Spring 1m \\
58787.587 & 16.62 $\pm$ 0.18 & u & Siding Spring 1m \\
58789.157 & 16.39 $\pm$ 0.17 & u & Swope \\
58790.943 & 16.09 $\pm$ 0.17 & u & Siding Spring 1m \\
58791.317 & 16.39 $\pm$ 0.17 & u & Swope \\
58793.521 & 15.95 $\pm$ 0.17 & u & Siding Spring 1m \\
58796.628 & 16.20 $\pm$ 0.17 & u & Siding Spring 1m \\
58802.499 & 16.75 $\pm$ 0.17 & u & Siding Spring 1m \\
58805.312 & 16.96 $\pm$ 0.17 & u & Swope \\
58805.571 & 16.43 $\pm$ 0.17 & u & Siding Spring 1m \\
58806.299 & 17.03 $\pm$ 0.17 & u & Swope \\
58807.301 & 16.97 $\pm$ 0.17 & u & Swope \\
58808.297 & 17.13 $\pm$ 0.17 & u & Swope \\
58808.589 & 16.67 $\pm$ 0.17 & u & Siding Spring 1m \\
58811.825 & 16.80 $\pm$ 0.17 & u & Siding Spring 1m \\
58813.183 & 17.16 $\pm$ 0.17 & u & Swope \\
58814.292 & 17.35 $\pm$ 0.17 & u & Swope \\
58814.634 & 16.99 $\pm$ 0.17 & u & Siding Spring 1m \\
58815.297 & 17.31 $\pm$ 0.17 & u & Swope \\
58817.241 & 17.33 $\pm$ 0.17 & u & Swope \\
58834.126 & 18.14 $\pm$ 0.17 & u & Swope \\
58835.257 & 18.19 $\pm$ 0.17 & u & Swope \\
58836.203 & 17.64 $\pm$ 0.17 & u & Swope \\
58839.207 & 17.95 $\pm$ 0.17 & u & Swope \\
58840.219 & 18.10 $\pm$ 0.17 & u & Swope \\
58849.225 & 18.57 $\pm$ 0.17 & u & Swope \\
58851.111 & 18.51 $\pm$ 0.17 & u & Swope \\
58853.088 & 18.69 $\pm$ 0.17 & u & Swope \\
58869.057 & 18.25 $\pm$ 0.17 & u & Swope \\
58871.096 & 19.27 $\pm$ 0.17 & u & Swope \\
58873.186 & 18.86 $\pm$ 0.17 & u & Swope \\
58882.185 & 19.18 $\pm$ 0.17 & u & Swope \\
58885.168 & 20.02 $\pm$ 0.17 & u & Swope \\
58893.082 & 19.45 $\pm$ 0.17 & u & Swope \\
58904.080 & 19.37 $\pm$ 0.17 & u & Swope \\
58909.081 & 20.77 $\pm$ 0.17 & u & Swope \\
58716.503 & $>$18.97 & g & Palomar 48in \\
58719.501 & $>$19.17 & g & Palomar 48in \\
58722.463 & $>$19.95 & g & Palomar 48in \\
58725.479 & $>$20.06 & g & Palomar 48in \\
58732.437 & $>$20.00 & g & Palomar 48in \\
58735.444 & $>$20.01 & g & Palomar 48in \\
58739.501 & $>$19.17 & g & Palomar 48in \\
58742.504 & $>$18.86 & g & Palomar 48in \\
58746.504 & 17.28 $\pm$ 0.05 & g & Palomar 48in \\
58748.521 & 16.90 $\pm$ 0.05 & g & Palomar 48in \\
58749.470 & 16.77 $\pm$ 0.04 & g & Palomar 48in \\
58751.495 & 16.57 $\pm$ 0.04 & g & Palomar 48in \\
58751.785 & 16.50 $\pm$ 0.01 & g & Siding Spring 1m \\
58752.348 & 16.52 $\pm$ 0.01 & g & Swope \\
58753.266 & 16.42 $\pm$ 0.01 & g & Swope \\
58754.331 & 16.38 $\pm$ 0.01 & g & Swope \\
58755.379 & 16.22 $\pm$ 0.01 & g & Swope \\
58756.423 & 16.11 $\pm$ 0.03 & g & Palomar 48in \\
58757.389 & 15.95 $\pm$ 0.04 & g & Palomar 48in \\
58757.946 & 15.84 $\pm$ 0.01 & g & Siding Spring 1m \\
58759.327 & 15.70 $\pm$ 0.01 & g & Swope \\
58759.439 & 15.70 $\pm$ 0.04 & g & Palomar 48in \\
58760.364 & 15.57 $\pm$ 0.01 & g & Swope \\
58760.936 & 15.49 $\pm$ 0.01 & g & Siding Spring 1m \\
58761.331 & 15.61 $\pm$ 0.01 & g & Swope \\
58762.261 & 15.50 $\pm$ 0.01 & g & Swope \\
58763.459 & 15.44 $\pm$ 0.02 & g & Palomar 48in \\
58763.931 & 15.36 $\pm$ 0.01 & g & Siding Spring 1m \\
58766.408 & 15.49 $\pm$ 0.03 & g & Palomar 48in \\
58766.771 & 15.49 $\pm$ 0.01 & g & Siding Spring 1m \\
58769.556 & 15.62 $\pm$ 0.02 & g & Siding Spring 1m \\
58772.336 & 15.67 $\pm$ 0.04 & g & Palomar 48in \\
58772.552 & 15.79 $\pm$ 0.02 & g & Siding Spring 1m \\
58772.565 & 15.75 $\pm$ 0.02 & g & Siding Spring 1m \\
58774.315 & 15.77 $\pm$ 0.01 & g & Swope \\
58775.274 & 15.80 $\pm$ 0.01 & g & Swope \\
58775.438 & 15.80 $\pm$ 0.04 & g & Palomar 48in \\
58776.321 & 15.82 $\pm$ 0.01 & g & Swope \\
58777.323 & 15.92 $\pm$ 0.01 & g & Swope \\
58778.241 & 16.01 $\pm$ 0.01 & g & Swope \\
58778.312 & 15.92 $\pm$ 0.04 & g & Palomar 48in \\
58778.580 & 15.92 $\pm$ 0.01 & g & Siding Spring 1m \\
58781.414 & 16.10 $\pm$ 0.04 & g & Palomar 48in \\
58781.732 & 16.08 $\pm$ 0.01 & g & Siding Spring 1m \\
58783.316 & 16.11 $\pm$ 0.01 & g & Swope \\
58784.267 & 16.27 $\pm$ 0.01 & g & Swope \\
58784.727 & 16.24 $\pm$ 0.01 & g & Siding Spring 1m \\
58785.284 & 16.25 $\pm$ 0.01 & g & Swope \\
58786.298 & 16.28 $\pm$ 0.01 & g & Swope \\
58787.247 & 16.31 $\pm$ 0.01 & g & Swope \\
58787.433 & 16.38 $\pm$ 0.05 & g & Palomar 48in \\
58787.689 & 16.36 $\pm$ 0.01 & g & Siding Spring 1m \\
58789.155 & 16.41 $\pm$ 0.01 & g & Swope \\
58790.335 & 16.49 $\pm$ 0.04 & g & Palomar 48in \\
58790.941 & 16.38 $\pm$ 0.01 & g & Siding Spring 1m \\
58791.316 & 16.53 $\pm$ 0.01 & g & Swope \\
58793.350 & 16.61 $\pm$ 0.04 & g & Palomar 48in \\
58793.504 & 16.63 $\pm$ 0.01 & g & Siding Spring 1m \\
58796.632 & 16.69 $\pm$ 0.01 & g & Siding Spring 1m \\
58799.833 & 16.89 $\pm$ 0.04 & g & Siding Spring 1m \\
58800.236 & 16.86 $\pm$ 0.07 & g & Palomar 48in \\
58802.508 & 16.97 $\pm$ 0.02 & g & Siding Spring 1m \\
58803.226 & 16.92 $\pm$ 0.14 & g & Palomar 48in \\
58804.295 & 17.15 $\pm$ 0.01 & g & Swope \\
58805.310 & 17.12 $\pm$ 0.01 & g & Swope \\
58805.518 & 17.04 $\pm$ 0.02 & g & Siding Spring 1m \\
58806.297 & 17.12 $\pm$ 0.01 & g & Swope \\
58806.372 & 17.10 $\pm$ 0.06 & g & Palomar 48in \\
58807.300 & 17.19 $\pm$ 0.01 & g & Swope \\
58808.296 & 17.16 $\pm$ 0.01 & g & Swope \\
58808.566 & 17.11 $\pm$ 0.02 & g & Siding Spring 1m \\
58811.817 & 17.22 $\pm$ 0.02 & g & Siding Spring 1m \\
58812.333 & 17.34 $\pm$ 0.05 & g & Palomar 48in \\
58813.182 & 17.32 $\pm$ 0.01 & g & Swope \\
58814.291 & 17.43 $\pm$ 0.01 & g & Swope \\
58814.524 & 17.39 $\pm$ 0.02 & g & Siding Spring 1m \\
58815.296 & 17.39 $\pm$ 0.01 & g & Swope \\
58817.240 & 17.53 $\pm$ 0.01 & g & Swope \\
58820.230 & 17.56 $\pm$ 0.13 & g & Palomar 48in \\
58829.239 & 17.98 $\pm$ 0.03 & g & Swope \\
58830.180 & 17.99 $\pm$ 0.02 & g & Swope \\
58831.187 & 17.98 $\pm$ 0.03 & g & Swope \\
58833.208 & 17.88 $\pm$ 0.10 & g & Palomar 48in \\
58834.122 & 18.00 $\pm$ 0.01 & g & Swope \\
58835.254 & 18.08 $\pm$ 0.01 & g & Swope \\
58836.202 & 18.15 $\pm$ 0.01 & g & Swope \\
58839.205 & 18.26 $\pm$ 0.01 & g & Swope \\
58840.217 & 18.24 $\pm$ 0.01 & g & Swope \\
58849.221 & 18.56 $\pm$ 0.01 & g & Swope \\
58851.106 & 18.52 $\pm$ 0.01 & g & Swope \\
58853.083 & 18.51 $\pm$ 0.02 & g & Swope \\
58863.148 & 19.07 $\pm$ 0.14 & g & Palomar 48in \\
58867.188 & 18.83 $\pm$ 0.12 & g & Palomar 48in \\
58869.054 & 18.87 $\pm$ 0.02 & g & Swope \\
58871.088 & 18.89 $\pm$ 0.02 & g & Swope \\
58872.194 & 19.02 $\pm$ 0.15 & g & Palomar 48in \\
58873.127 & 18.95 $\pm$ 0.10 & g & Palomar 48in \\
58873.182 & 18.97 $\pm$ 0.01 & g & Swope \\
58874.188 & 19.12 $\pm$ 0.17 & g & Palomar 48in \\
58875.166 & 19.09 $\pm$ 0.11 & g & Palomar 48in \\
58876.147 & 19.04 $\pm$ 0.01 & g & Swope \\
58876.169 & 19.01 $\pm$ 0.12 & g & Palomar 48in \\
58876.169 & 19.01 $\pm$ 0.11 & g & Palomar 48in \\
58877.169 & 19.16 $\pm$ 0.17 & g & Palomar 48in \\
58877.178 & 19.09 $\pm$ 0.02 & g & Swope \\
58878.213 & 19.14 $\pm$ 0.26 & g & Palomar 48in \\
58880.126 & 19.10 $\pm$ 0.16 & g & Palomar 48in \\
58881.147 & 19.18 $\pm$ 0.20 & g & Palomar 48in \\
58881.148 & 19.03 $\pm$ 0.17 & g & Palomar 48in \\
58882.179 & 19.11 $\pm$ 0.03 & g & Swope \\
58885.125 & 19.08 $\pm$ 0.22 & g & Palomar 48in \\
58885.164 & 19.37 $\pm$ 0.09 & g & Swope \\
58891.186 & 19.32 $\pm$ 0.18 & g & Palomar 48in \\
58893.078 & 19.32 $\pm$ 0.01 & g & Swope \\
58893.168 & 19.53 $\pm$ 0.17 & g & Palomar 48in \\
58894.171 & 19.24 $\pm$ 0.14 & g & Palomar 48in \\
58894.172 & 19.33 $\pm$ 0.19 & g & Palomar 48in \\
58895.144 & 19.33 $\pm$ 0.19 & g & Palomar 48in \\
58896.148 & 19.43 $\pm$ 0.14 & g & Swope \\
58897.120 & 19.37 $\pm$ 0.02 & g & Swope \\
58898.171 & 19.46 $\pm$ 0.20 & g & Palomar 48in \\
58898.179 & 19.56 $\pm$ 0.20 & g & Palomar 48in \\
58899.149 & 19.58 $\pm$ 0.16 & g & Palomar 48in \\
58904.075 & 19.53 $\pm$ 0.02 & g & Swope \\
58904.133 & 19.65 $\pm$ 0.20 & g & Palomar 48in \\
58909.076 & 19.49 $\pm$ 0.02 & g & Swope \\
58911.110 & 19.61 $\pm$ 0.04 & g & Swope \\
58921.073 & 19.71 $\pm$ 0.04 & g & Swope \\
58716.481 & $>$19.30 & r & Palomar 48in \\
58719.477 & $>$19.16 & r & Palomar 48in \\
58722.500 & $>$20.20 & r & Palomar 48in \\
58725.497 & $>$20.26 & r & Palomar 48in \\
58730.479 & $>$19.86 & r & Palomar 48in \\
58735.487 & $>$20.10 & r & Palomar 48in \\
58739.462 & $>$19.42 & r & Palomar 48in \\
58745.500 & $>$19.10 & r & Palomar 48in \\
58745.500 & 17.57 $\pm$ 0.07 & r & Palomar 48in \\
58748.481 & 17.18 $\pm$ 0.05 & r & Palomar 48in \\
58751.502 & 16.75 $\pm$ 0.04 & r & Palomar 48in \\
58751.783 & 16.74 $\pm$ 0.01 & r & Siding Spring 1m \\
58752.344 & 16.67 $\pm$ 0.01 & r & Swope \\
58753.261 & 16.62 $\pm$ 0.01 & r & Swope \\
58754.327 & 16.56 $\pm$ 0.01 & r & Swope \\
58754.762 & 16.49 $\pm$ 0.01 & r & Siding Spring 1m \\
58755.376 & 16.36 $\pm$ 0.01 & r & Swope \\
58757.474 & 16.10 $\pm$ 0.04 & r & Palomar 48in \\
58757.944 & 16.07 $\pm$ 0.01 & r & Siding Spring 1m \\
58759.323 & 15.87 $\pm$ 0.01 & r & Swope \\
58760.361 & 15.76 $\pm$ 0.00 & r & Swope \\
58760.947 & 15.76 $\pm$ 0.01 & r & Siding Spring 1m \\
58761.328 & 15.70 $\pm$ 0.00 & r & Swope \\
58762.259 & 15.65 $\pm$ 0.00 & r & Swope \\
58763.499 & 15.57 $\pm$ 0.03 & r & Palomar 48in \\
58763.928 & 15.59 $\pm$ 0.01 & r & Siding Spring 1m \\
58763.945 & 15.57 $\pm$ 0.01 & r & Siding Spring 1m \\
58766.482 & 15.58 $\pm$ 0.03 & r & Palomar 48in \\
58766.603 & 15.70 $\pm$ 0.13 & r & Siding Spring 1m \\
58766.767 & 15.65 $\pm$ 0.01 & r & Siding Spring 1m \\
58769.431 & 15.67 $\pm$ 0.04 & r & Palomar 48in \\
58769.559 & 15.75 $\pm$ 0.03 & r & Siding Spring 1m \\
58772.379 & 15.78 $\pm$ 0.03 & r & Palomar 48in \\
58772.548 & 15.89 $\pm$ 0.03 & r & Siding Spring 1m \\
58774.312 & 15.83 $\pm$ 0.01 & r & Swope \\
58775.357 & 15.81 $\pm$ 0.03 & r & Palomar 48in \\
58776.318 & 15.89 $\pm$ 0.01 & r & Swope \\
58777.321 & 15.92 $\pm$ 0.01 & r & Swope \\
58778.239 & 16.17 $\pm$ 0.01 & r & Swope \\
58778.458 & 15.91 $\pm$ 0.03 & r & Palomar 48in \\
58778.574 & 16.00 $\pm$ 0.01 & r & Siding Spring 1m \\
58781.958 & 16.11 $\pm$ 0.01 & r & Siding Spring 1m \\
58783.312 & 16.12 $\pm$ 0.01 & r & Swope \\
58784.264 & 16.24 $\pm$ 0.00 & r & Swope \\
58784.392 & 16.18 $\pm$ 0.05 & r & Palomar 48in \\
58784.729 & 16.26 $\pm$ 0.01 & r & Siding Spring 1m \\
58785.283 & 16.23 $\pm$ 0.01 & r & Swope \\
58786.295 & 16.30 $\pm$ 0.01 & r & Swope \\
58787.244 & 16.33 $\pm$ 0.01 & r & Swope \\
58787.583 & 16.38 $\pm$ 0.02 & r & Siding Spring 1m \\
58789.152 & 16.40 $\pm$ 0.01 & r & Swope \\
58790.369 & 16.44 $\pm$ 0.04 & r & Palomar 48in \\
58790.938 & 16.47 $\pm$ 0.01 & r & Siding Spring 1m \\
58791.314 & 16.49 $\pm$ 0.01 & r & Swope \\
58793.472 & 16.58 $\pm$ 0.04 & r & Palomar 48in \\
58793.506 & 16.60 $\pm$ 0.01 & r & Siding Spring 1m \\
58796.319 & 16.69 $\pm$ 0.05 & r & Palomar 48in \\
58796.625 & 16.70 $\pm$ 0.01 & r & Siding Spring 1m \\
58799.837 & 16.72 $\pm$ 0.03 & r & Siding Spring 1m \\
58802.505 & 16.91 $\pm$ 0.02 & r & Siding Spring 1m \\
58803.314 & 16.94 $\pm$ 0.04 & r & Palomar 48in \\
58804.293 & 16.92 $\pm$ 0.01 & r & Swope \\
58805.308 & 16.96 $\pm$ 0.01 & r & Swope \\
58805.516 & 17.04 $\pm$ 0.02 & r & Siding Spring 1m \\
58806.294 & 17.06 $\pm$ 0.01 & r & Swope \\
58806.317 & 17.06 $\pm$ 0.05 & r & Palomar 48in \\
58807.297 & 17.05 $\pm$ 0.01 & r & Swope \\
58808.294 & 17.17 $\pm$ 0.01 & r & Swope \\
58808.593 & 17.08 $\pm$ 0.03 & r & Siding Spring 1m \\
58811.815 & 17.19 $\pm$ 0.01 & r & Siding Spring 1m \\
58812.274 & 17.28 $\pm$ 0.06 & r & Palomar 48in \\
58813.179 & 17.24 $\pm$ 0.01 & r & Swope \\
58814.288 & 17.35 $\pm$ 0.01 & r & Swope \\
58814.549 & 17.38 $\pm$ 0.02 & r & Siding Spring 1m \\
58815.293 & 17.32 $\pm$ 0.01 & r & Swope \\
58817.238 & 17.48 $\pm$ 0.01 & r & Swope \\
58827.194 & 17.96 $\pm$ 0.02 & r & Swope \\
58828.207 & 18.03 $\pm$ 0.02 & r & Swope \\
58829.234 & 17.84 $\pm$ 0.03 & r & Swope \\
58830.175 & 18.20 $\pm$ 0.02 & r & Swope \\
58830.251 & 17.80 $\pm$ 0.09 & r & Palomar 48in \\
58831.183 & 17.83 $\pm$ 0.02 & r & Swope \\
58833.251 & 17.99 $\pm$ 0.07 & r & Palomar 48in \\
58833.325 & 17.99 $\pm$ 0.09 & r & Palomar 48in \\
58834.116 & 17.90 $\pm$ 0.01 & r & Swope \\
58835.250 & 17.89 $\pm$ 0.01 & r & Swope \\
58836.198 & 17.93 $\pm$ 0.01 & r & Swope \\
58837.237 & 18.00 $\pm$ 0.06 & r & Palomar 48in \\
58839.201 & 18.10 $\pm$ 0.01 & r & Swope \\
58840.213 & 18.10 $\pm$ 0.01 & r & Swope \\
58849.215 & 18.39 $\pm$ 0.01 & r & Swope \\
58851.098 & 18.39 $\pm$ 0.01 & r & Swope \\
58853.074 & 18.45 $\pm$ 0.02 & r & Swope \\
58854.169 & 18.45 $\pm$ 0.21 & r & Palomar 48in \\
58860.168 & 18.50 $\pm$ 0.10 & r & Palomar 48in \\
58860.208 & 18.55 $\pm$ 0.18 & r & Palomar 48in \\
58861.214 & 18.68 $\pm$ 0.15 & r & Palomar 48in \\
58867.231 & 18.75 $\pm$ 0.11 & r & Palomar 48in \\
58869.047 & 18.67 $\pm$ 0.02 & r & Swope \\
58871.084 & 18.78 $\pm$ 0.03 & r & Swope \\
58872.260 & 18.78 $\pm$ 0.14 & r & Palomar 48in \\
58873.161 & 18.94 $\pm$ 0.12 & r & Palomar 48in \\
58873.174 & 18.85 $\pm$ 0.02 & r & Swope \\
58875.196 & 18.99 $\pm$ 0.12 & r & Palomar 48in \\
58876.139 & 18.96 $\pm$ 0.01 & r & Swope \\
58877.146 & 19.08 $\pm$ 0.12 & r & Palomar 48in \\
58877.170 & 18.92 $\pm$ 0.02 & r & Swope \\
58878.170 & 18.94 $\pm$ 0.26 & r & Palomar 48in \\
58878.171 & 18.38 $\pm$ 0.17 & r & Palomar 48in \\
58881.107 & 18.92 $\pm$ 0.15 & r & Palomar 48in \\
58881.108 & 18.95 $\pm$ 0.16 & r & Palomar 48in \\
58882.169 & 19.82 $\pm$ 0.26 & r & Swope \\
58882.170 & 18.80 $\pm$ 0.02 & r & Swope \\
58885.155 & 18.95 $\pm$ 0.06 & r & Swope \\
58886.130 & 18.86 $\pm$ 0.17 & r & Palomar 48in \\
58887.126 & 18.90 $\pm$ 0.21 & r & Palomar 48in \\
58891.129 & 19.39 $\pm$ 0.19 & r & Palomar 48in \\
58893.069 & 19.77 $\pm$ 0.10 & r & Swope \\
58893.070 & 19.08 $\pm$ 0.02 & r & Swope \\
58894.144 & 19.26 $\pm$ 0.12 & r & Palomar 48in \\
58895.108 & 19.22 $\pm$ 0.12 & r & Palomar 48in \\
58896.122 & 19.15 $\pm$ 0.05 & r & Swope \\
58896.125 & 19.26 $\pm$ 0.04 & r & Swope \\
58897.109 & 19.30 $\pm$ 0.09 & r & Swope \\
58897.110 & 19.03 $\pm$ 0.02 & r & Swope \\
58898.129 & 19.27 $\pm$ 0.14 & r & Palomar 48in \\
58904.065 & 19.82 $\pm$ 0.13 & r & Swope \\
58904.066 & 19.25 $\pm$ 0.02 & r & Swope \\
58909.065 & 19.23 $\pm$ 0.02 & r & Swope \\
58911.099 & 19.45 $\pm$ 0.03 & r & Swope \\
58921.062 & 19.34 $\pm$ 0.03 & r & Swope \\
58752.346 & 16.86 $\pm$ 0.01 & i & Swope \\
58753.264 & 16.79 $\pm$ 0.01 & i & Swope \\
58754.329 & 16.67 $\pm$ 0.01 & i & Swope \\
58755.378 & 16.66 $\pm$ 0.01 & i & Swope \\
58759.325 & 16.07 $\pm$ 0.01 & i & Swope \\
58760.363 & 15.92 $\pm$ 0.01 & i & Swope \\
58761.330 & 15.88 $\pm$ 0.00 & i & Swope \\
58762.260 & 15.84 $\pm$ 0.01 & i & Swope \\
58774.314 & 15.95 $\pm$ 0.01 & i & Swope \\
58776.320 & 16.01 $\pm$ 0.01 & i & Swope \\
58777.322 & 16.08 $\pm$ 0.01 & i & Swope \\
58778.240 & 16.29 $\pm$ 0.01 & i & Swope \\
58783.314 & 16.24 $\pm$ 0.01 & i & Swope \\
58784.265 & 16.26 $\pm$ 0.01 & i & Swope \\
58785.286 & 16.37 $\pm$ 0.01 & i & Swope \\
58786.297 & 16.42 $\pm$ 0.01 & i & Swope \\
58787.246 & 16.49 $\pm$ 0.01 & i & Swope \\
58787.248 & 16.46 $\pm$ 0.01 & i & Swope \\
58789.153 & 16.77 $\pm$ 0.01 & i & Swope \\
58791.315 & 16.67 $\pm$ 0.01 & i & Swope \\
58804.294 & 17.19 $\pm$ 0.01 & i & Swope \\
58805.309 & 17.26 $\pm$ 0.01 & i & Swope \\
58806.296 & 17.24 $\pm$ 0.02 & i & Swope \\
58807.299 & 17.27 $\pm$ 0.01 & i & Swope \\
58808.295 & 17.23 $\pm$ 0.02 & i & Swope \\
58813.181 & 17.48 $\pm$ 0.01 & i & Swope \\
58814.290 & 17.52 $\pm$ 0.01 & i & Swope \\
58815.295 & 17.54 $\pm$ 0.02 & i & Swope \\
58817.239 & 17.76 $\pm$ 0.01 & i & Swope \\
58827.196 & 18.14 $\pm$ 0.02 & i & Swope \\
58828.209 & 18.11 $\pm$ 0.02 & i & Swope \\
58829.237 & 18.04 $\pm$ 0.03 & i & Swope \\
58830.177 & 18.22 $\pm$ 0.02 & i & Swope \\
58831.185 & 18.10 $\pm$ 0.03 & i & Swope \\
58834.119 & 18.16 $\pm$ 0.02 & i & Swope \\
58835.252 & 18.18 $\pm$ 0.02 & i & Swope \\
58836.200 & 18.27 $\pm$ 0.01 & i & Swope \\
58839.204 & 18.24 $\pm$ 0.01 & i & Swope \\
58840.215 & 18.29 $\pm$ 0.02 & i & Swope \\
58849.218 & 18.67 $\pm$ 0.02 & i & Swope \\
58851.102 & 18.58 $\pm$ 0.01 & i & Swope \\
58853.078 & 18.73 $\pm$ 0.03 & i & Swope \\
58869.051 & 18.92 $\pm$ 0.04 & i & Swope \\
58871.092 & 18.97 $\pm$ 0.05 & i & Swope \\
58873.178 & 19.24 $\pm$ 0.04 & i & Swope \\
58876.143 & 19.35 $\pm$ 0.02 & i & Swope \\
58877.174 & 19.16 $\pm$ 0.03 & i & Swope \\
58882.174 & 19.22 $\pm$ 0.04 & i & Swope \\
58885.159 & 19.30 $\pm$ 0.08 & i & Swope \\
58893.074 & 19.38 $\pm$ 0.03 & i & Swope \\
58896.143 & 19.84 $\pm$ 0.14 & i & Swope \\
58897.115 & 19.61 $\pm$ 0.04 & i & Swope \\
58904.070 & 19.67 $\pm$ 0.03 & i & Swope \\
58909.070 & 19.55 $\pm$ 0.04 & i & Swope \\
58911.104 & 19.47 $\pm$ 0.06 & i & Swope \\
58921.068 & 19.78 $\pm$ 0.07 & i & Swope